# Long Pulse Modulators

*H-J. Eckoldt*
DESY, Hamburg, Germany

**Abstract**
Long pulse modulators are used to produce high-voltage, high-power pulses with durations of several hundred microseconds up to some milliseconds. The loads are one or more klystrons for producing RF power to accelerate the particle beam in superconducting cavities. After years of development and improvements in different institutes a variety of topologies exist, and are presented. The basics of modulators, pulse requirements and klystrons are explained. Additionally, the charging of internal energy storage will be addressed. The outlook for future developments is given.

## 1 Introduction

The development of long pulse modulators started in the early 1990s in combination with the development of superconducting cavities for accelerating particles. The design aim at that time was a linear accelerator called TESLA, which would have had a length of approximately 30 km. Although this accelerator was not built, the superconducting technology for cavities is being implemented in the European X-ray Free-Electron Laser Facility (XFEL) that is currently under construction in Hamburg.

Operating such an accelerator in continuous mode would form a large heat load for the cryogenics. The operating cost would be extremely high. Therefore these cavities are operated in pulsed mode with a duty factor of approximately 1.5%. As a consequence the high voltage supply to the klystron is pulsed for the duration of the RF that is required.

Long pulse modulators produce pulsed high voltage with a pulse length of a few 100 µs to some milliseconds. Fermilab National Accelerator Laboratory (Fermilab) [1] developed the first modulators of this type, which were later operated at DESY (Fig. 1). Here, further improvements and technology transfer to manufacturers were carried out.

In the following the basics of the RF station with the main components, basic specifications of the pulse and different types of modulators will be explained, including examples of modulators that have been built and operated in different laboratories. Finally, the outlook for future developments will be given.

## 2 XFEL RF station

The modulator is part of an RF station. Figure 2 shows the topology of the RF station at XFEL [2]. The unit producing the pulsed power is called a modulator. The high-voltage power supply (HVPS) takes electrical power from the grid and stores energy in order not to disturb the grid. By means of the pulse-forming unit high-voltage pulses are generated at a voltage level of 10 kV. A pulse transformer transforms the pulses to 120 kV. This is the voltage that is required by the klystron in which RF power is produced. Waveguides transport the RF to the superconducting cavities.

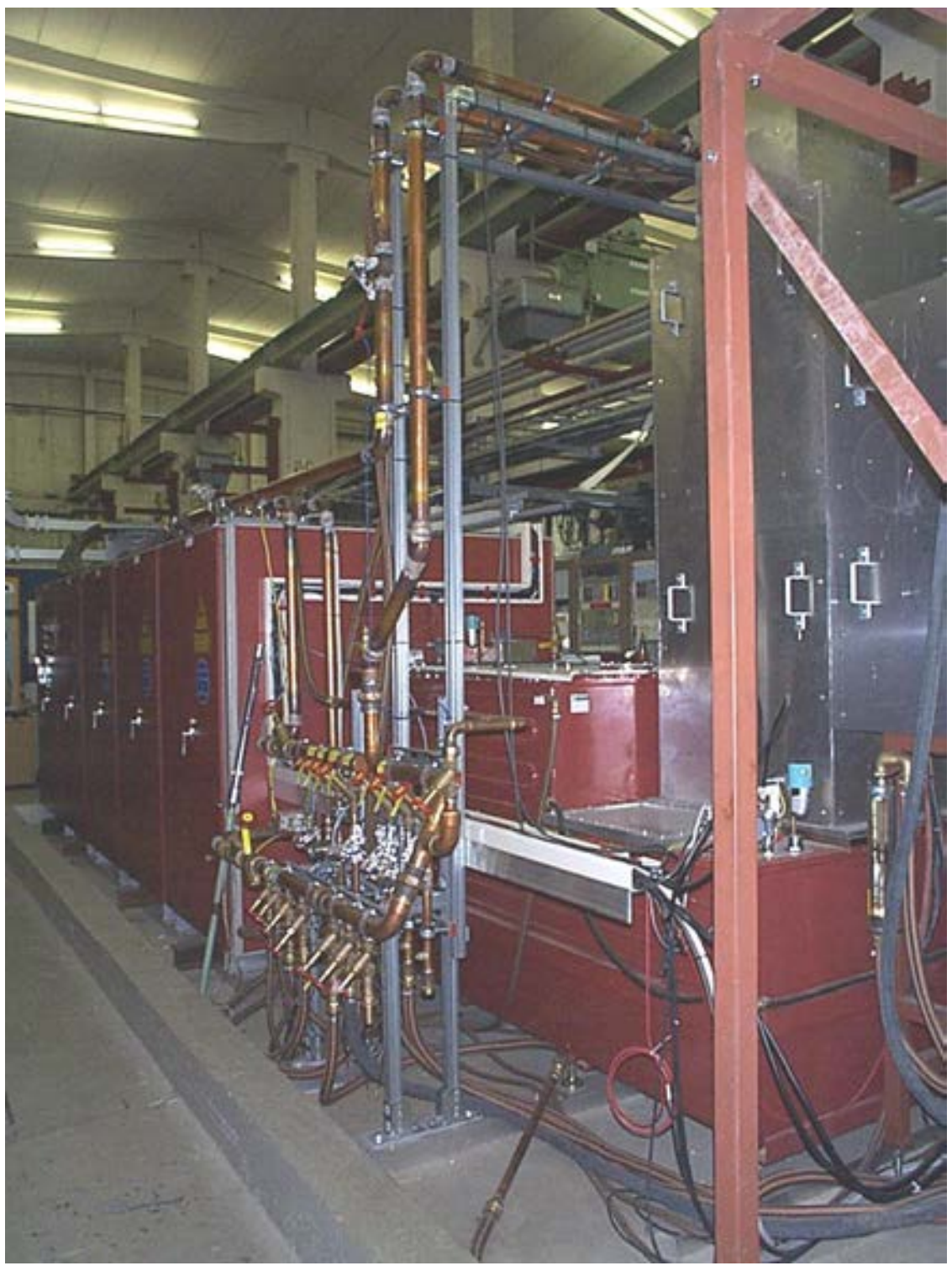

**Fig. 1:** Bouncer modulator developed at FNAL

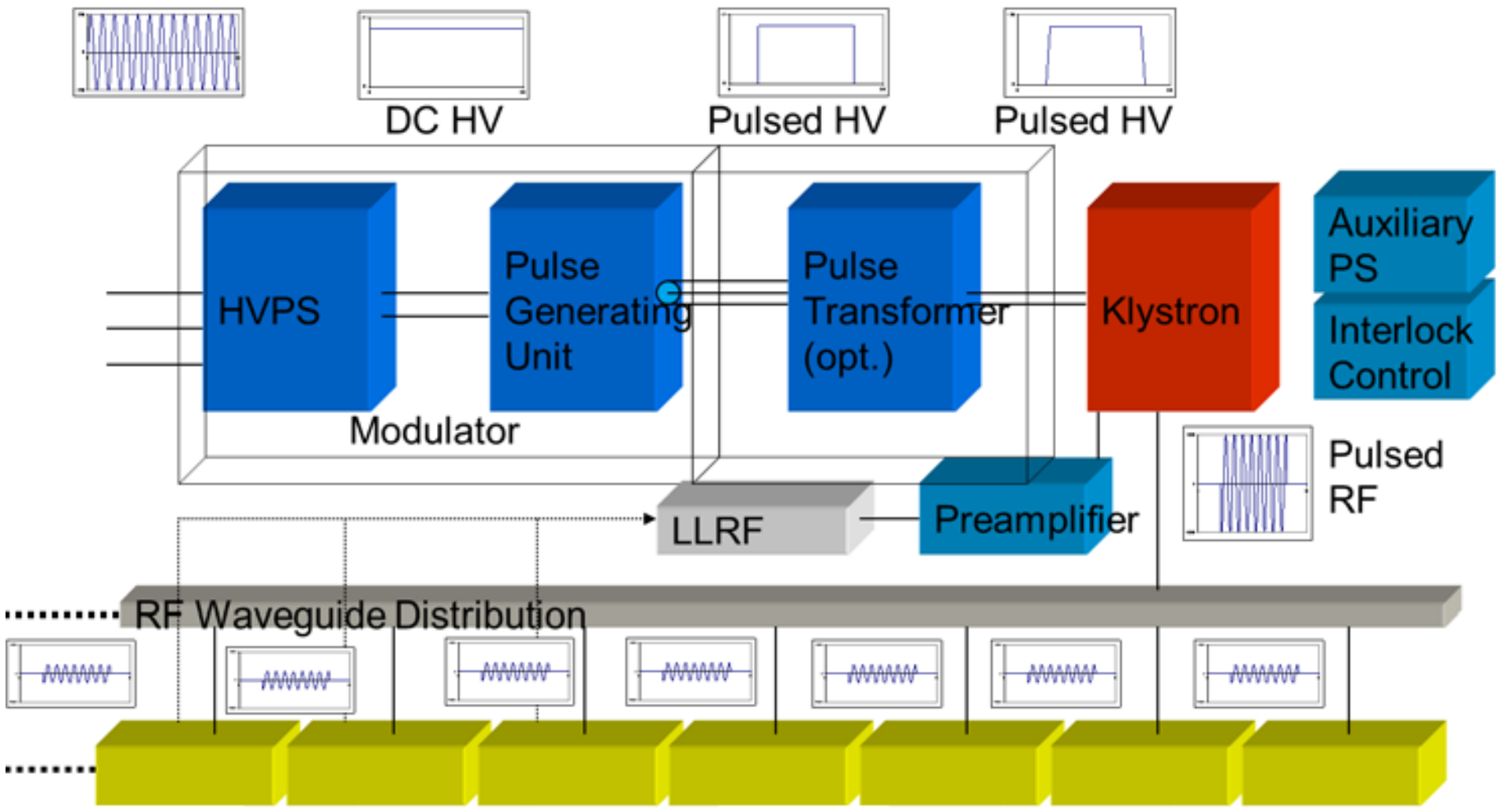

**Fig. 2:** Typical RF station as used in XFEL

## 2.1 Klystron

The klystron is a vacuum tube in which RF power is generated. A low-power RF signal is amplified and high-power RF is provided. A klystron is operated at high voltages, for example the nominal XFEL klystron voltage is 115 kV. At this working point, it can generate 10 MW RF at 1.3 GHz for 1.34 ms, which is 200 µs less than the modulator pulse. For the modulator the klystron is a load with the following relation between current and klystron voltage:

$$I = \mu P \times U^{3/2} \tag{1}$$

where $I$ is klystron current, $U$ is klystron voltage and $\mu P$ is µPerveance.

The μPerveance is a parameter of the klystron. It is determined by the geometry and is constant for each klystron. The characteristic curves of the klystron represented by Eq. (1) are given in Figs. 3 and 4. These are the klystron current as a function of voltage (Fig. 3) and the resistance at the operating point as a function of voltage (Fig. 4).

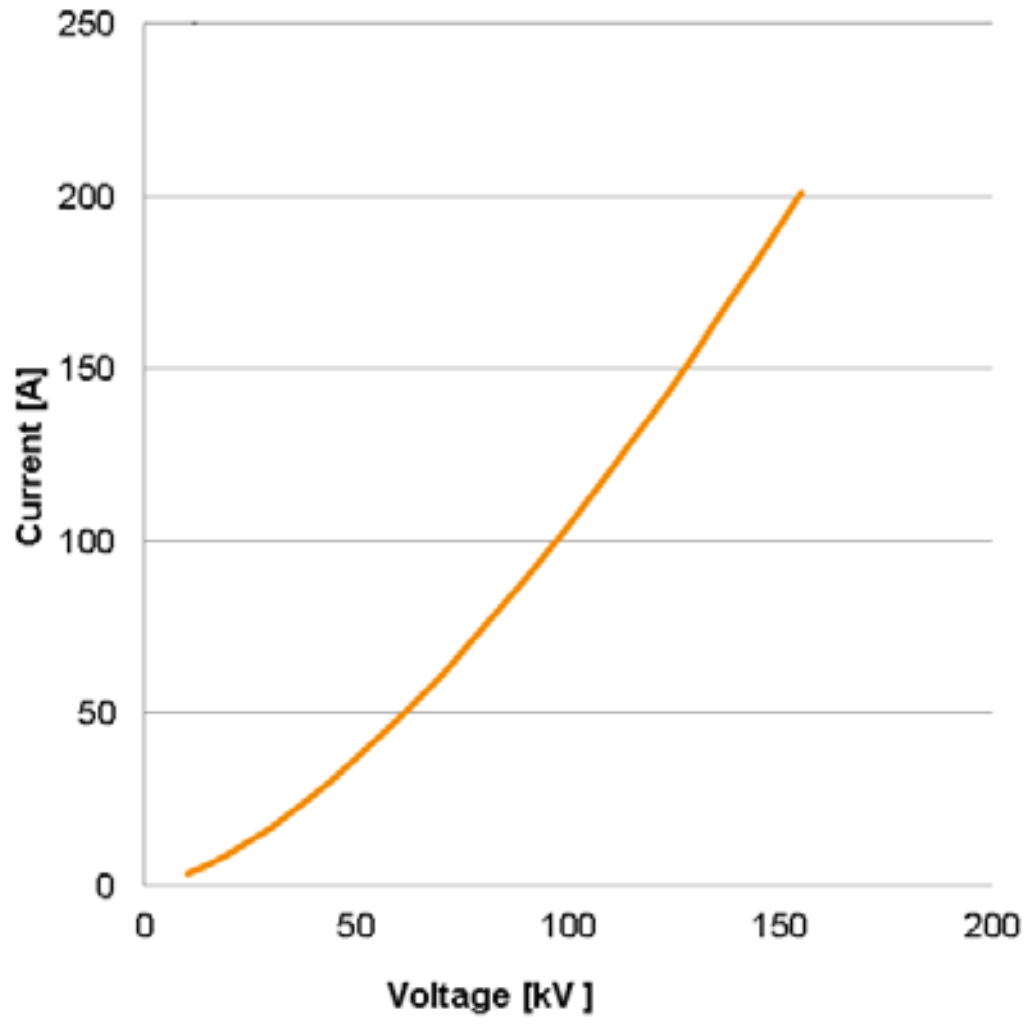

**Fig. 3:** Klystron current as $f(U_{\mathrm{klystron}})$

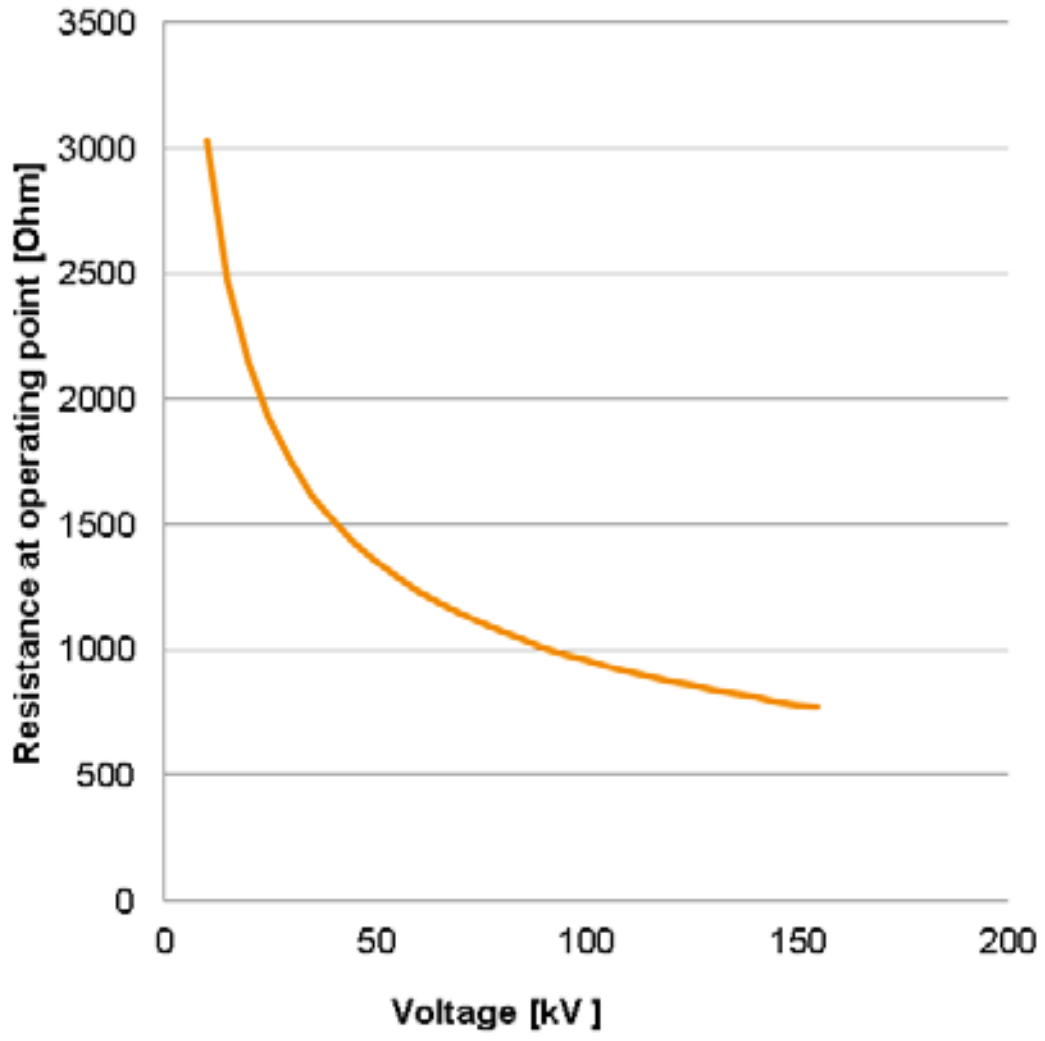

**Fig. 4:** Equivalent resistance as $f(U_{\mathrm{klystron}})$

For simulation and calculation of the modulator behaviour it is possible to use a simple resistor with series diode. To enhance simulations in a wide range of operation, the resistance has to vary according to the characteristic curve.

The beam power of the klystron is approximately

$$P_{\mathrm{Beam}} = \mu P \times U^{5/2} \quad (2)$$

and RF power is

$$P_{\mathrm{RF}} = \eta P_{\mathrm{Beam}} \quad (3)$$

where $\eta$ is the efficiency of the klystron.

### 2.2 Electrical data for the XFEL Thales TH801 multibeam klystron

Figure 5 shows the Thales TH801 multibeam klystron used in the XFEL, and electrical data are given in Table 1.

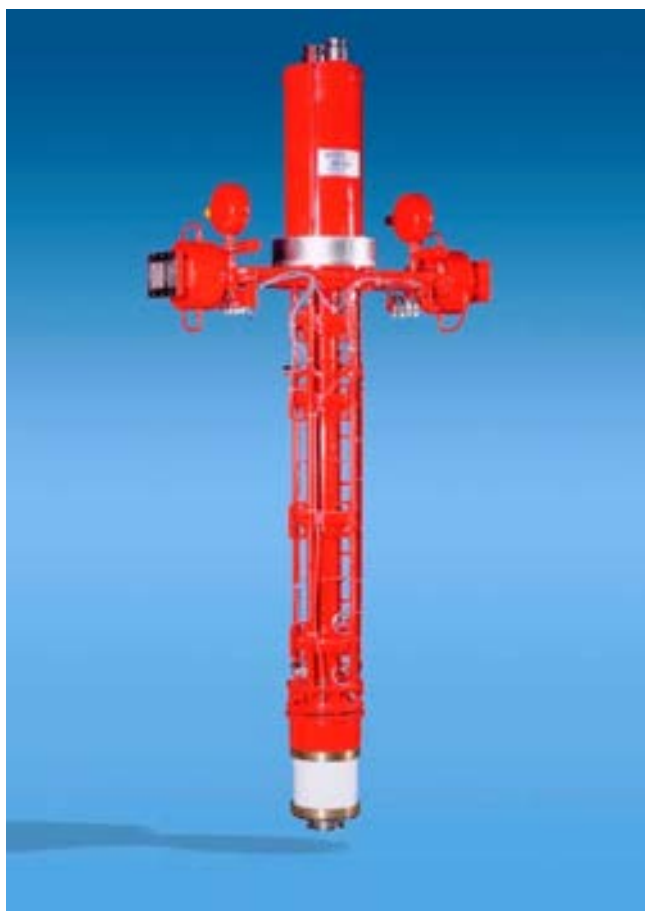

**Fig. 5:** Thales TH801 multibeam klystrom

**Table 1:** Electrical data for the Thales TH801 multibeam klystron

| Parameter | Value |
|---|---|
| Cathode voltage | 117 kV |
| Beam current | 131 A |
| µPerveance | 3.27 |
| Electrical resistance | 893 Ω at 117 kV |
| Maximum RF peak power | 10 MW |
| Electrical power | 15.33 MW |
| Electrical pulse duration | 1.54 ms (1.7 ms max) |
| RF pulse duration | 1.34 ms |
| Repetition rate | 10 Hz |
| Efficiency | 65% |
| RF average power | 150 kW |
| Average electrical power | 280 kW |

### 2.3 Klystron arcing

During operation arcs inside the klystron may occasionally occur. The HV collapses to the burning voltage of the arc and the short-circuit current increases. When an arc occurs the modulator has to detect it, interrupt the energy supply to the klystron and dissipate all energy stored in the current path to the klystron, in chokes, capacitors or even long cables, for example. Only 10 J to 20 J are allowed to be deposited in the klystron. More would damage the surface, leading to a derating or even damage to the entire klystron.

The equivalent circuit for an arcing klystron is a series combination of a voltage source of 100 V and a resistor. The 100 V corresponds to the burning voltage of the arc. Actually, no real measurements are available. In the literature, values of arc voltages are between 30 V to 100 V. To be on the safe side, for protection of the klystron, the conservative value was taken. The resistor represents a current-dependent component. The resistance is assumed to be 100 mΩ.

The equivalent circuit for the entire klystron is shown in Fig. 6. It represents the characteristic resistance given in Fig. 4 and the arc-model circuit. The following simulations are based on this equivalent circuit. During normal operation only the characteristic curve is valid, but it is important to have the protective part for the fault in mind.

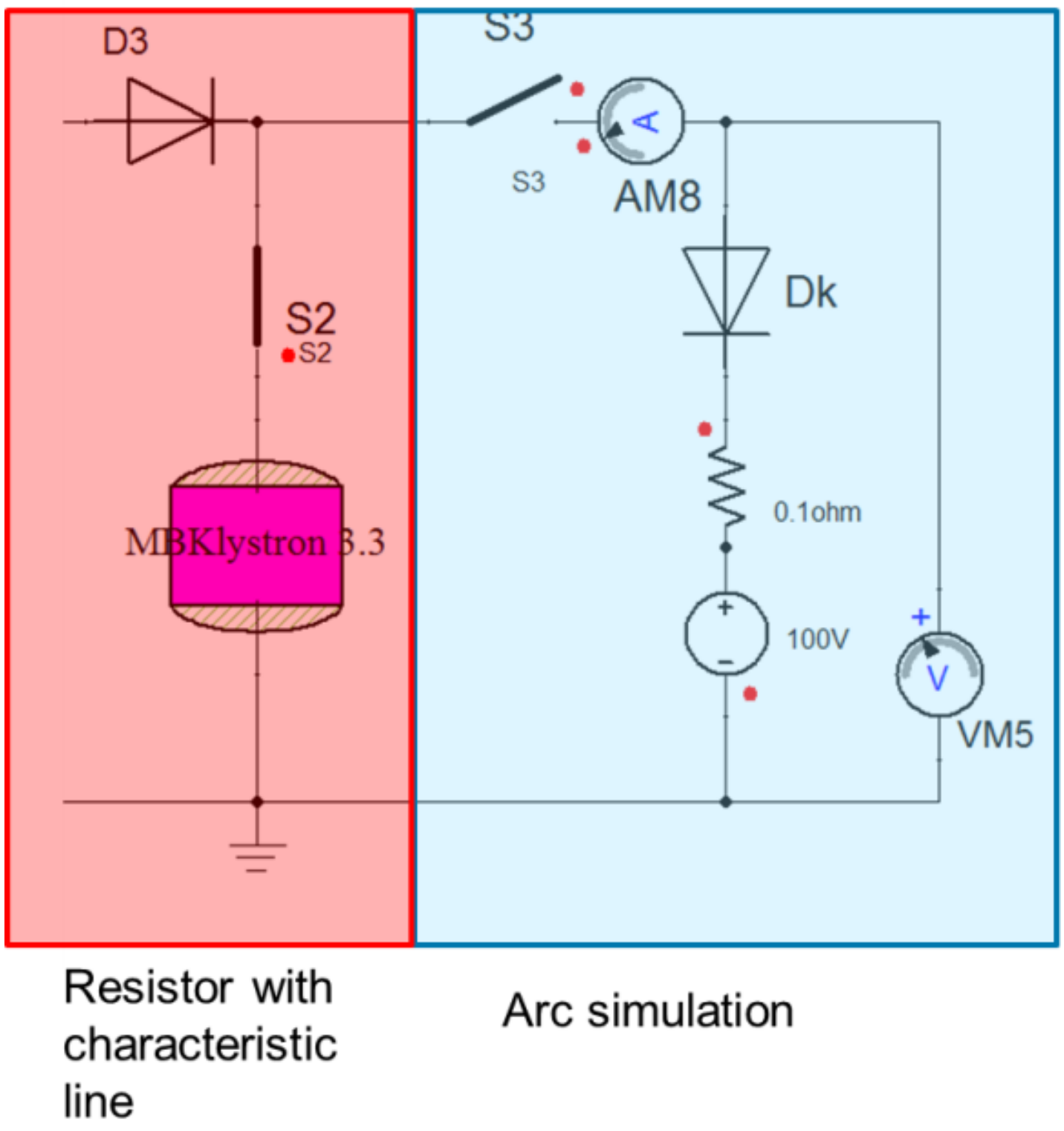

**Fig. 6:** Simulation model of a klystron

## 3 Definition of the pulse

When talking about pulsed power a few definitions are given to describe the performance of modulators and the pulse. These might differ at different laboratories.

**Table 2:** Pulsed power definitions

| **Term** | **Definition** |
|---|---|
| HV pulse | At DESY the HV pulse duration is given from turn on of the pulse-forming unit, other labs use an uptime of, for example, 70% of the HV. |
| Rise time | Time from the start up to the flat top. This is often defined as 10% to 90% or 10% to 99%. |
| Flat top | The time when the pulse is at the klystron’s operating voltage. Variations, called ripple, lead to RF phase shifts that have to be compensated for by the Low Level RF system (LLRF). The flat top is defined as *x*% or ±*x*% of the voltage. At XFEL 0.3% pp was defined. |
| Fall time | The time that the modulator voltage requires to drop down. |
| Reverse voltage | Also known as the undershoot, this is the maximum allowed negative voltage, it is approximately 20% of the klystron operating voltage |
| Repetition rate | Frequency of pulse repetition. |
| Pulse to pulse stability | Value of the repeating value of the flat top. |

Figures 7 and 8 show the above-defined values.

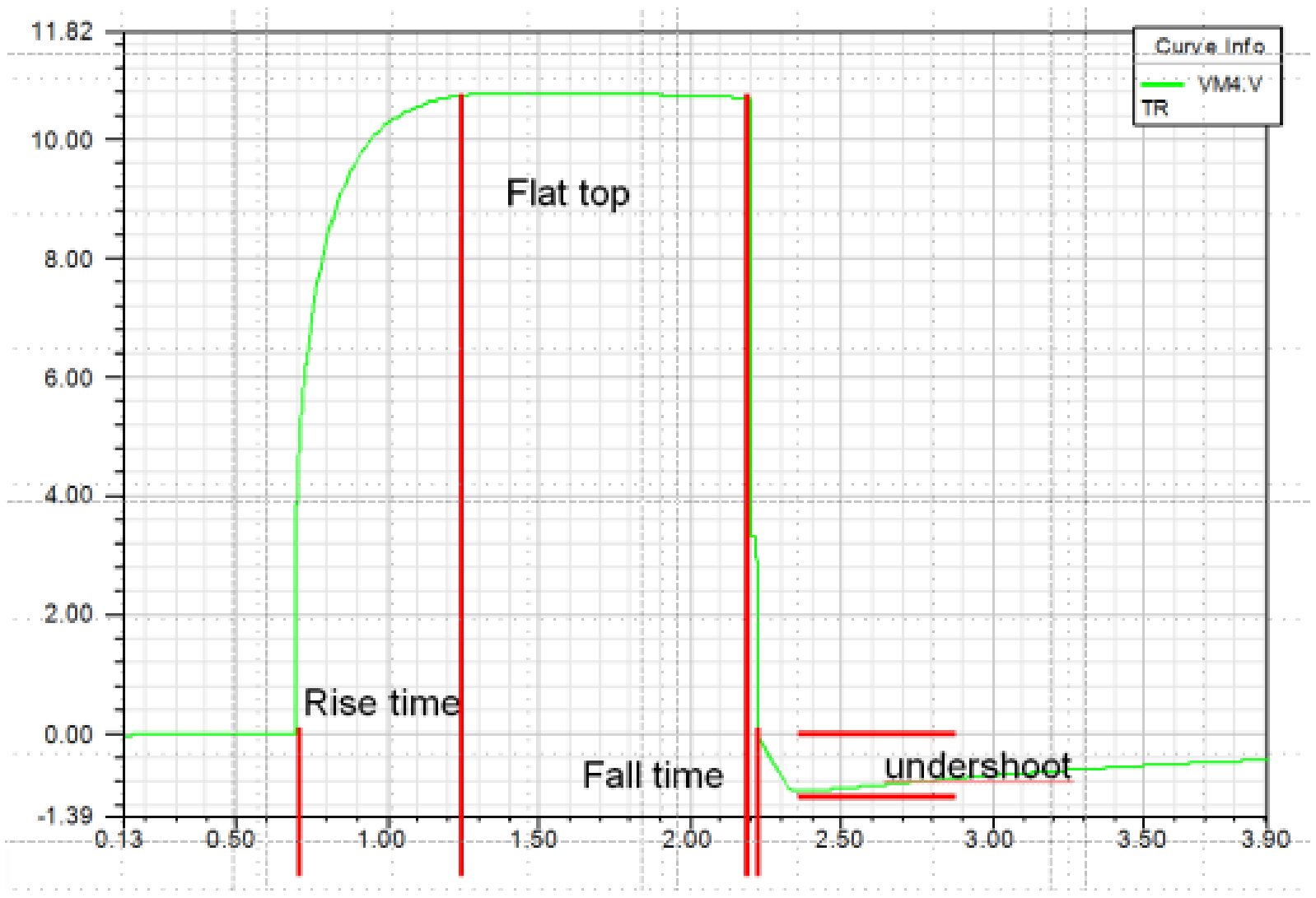

**Fig. 7:** Definitions of the output pulse

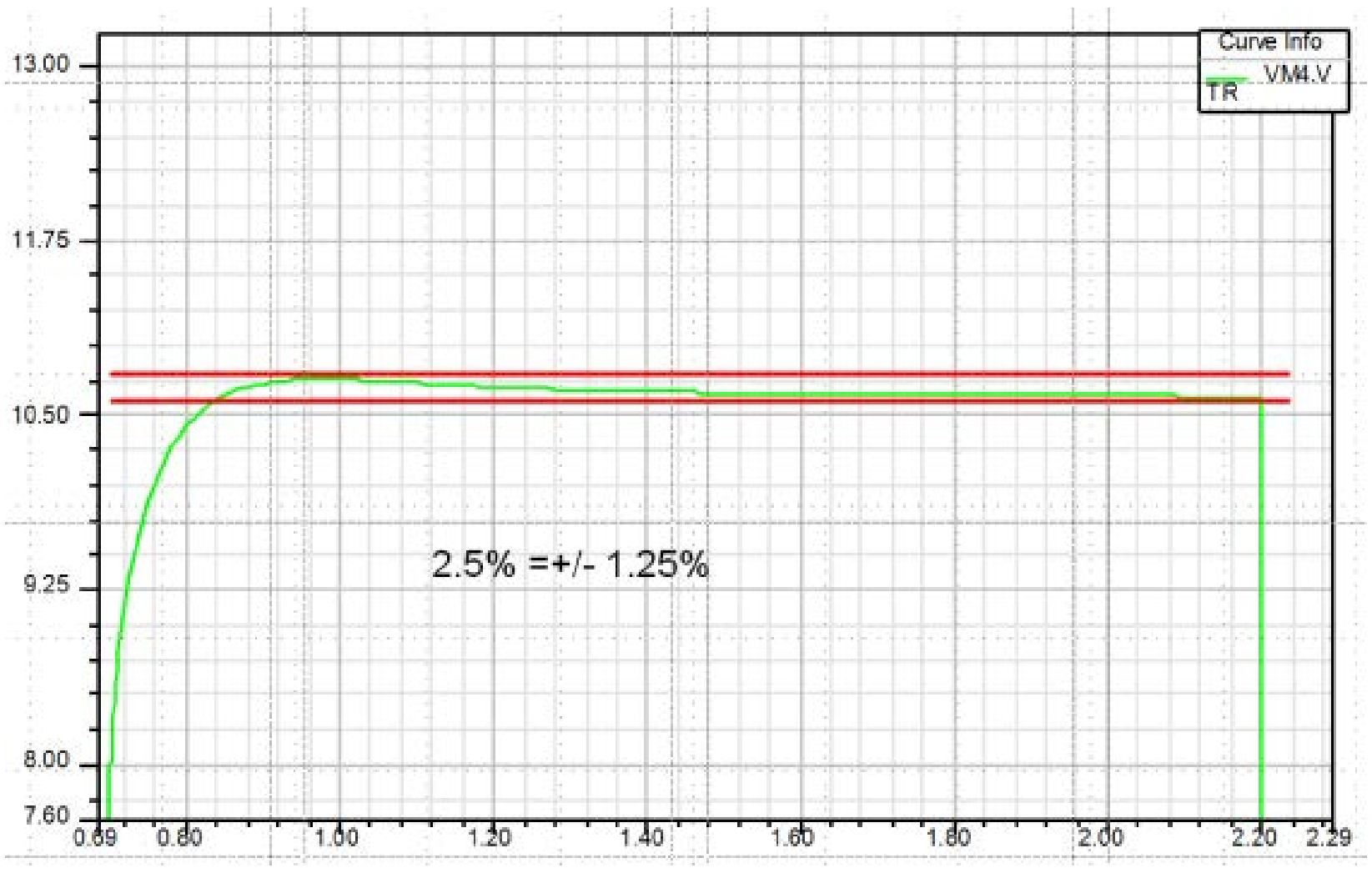

**Fig. 8:** Definition of flat top

## 4 Modulator basics

There are different modulator topologies. To understand the principle of modulators, in the following sections a bouncer modulator will be built step-by-step and real examples will be given. For the calculations the klystron data and the requirements for the XFEL are taken. The design pulse length for the hardware was chosen to be 1.7 ms.

### 4.1 Direct switching

A modulator contains energy storage. From this storage the energy is supplied to the klystron. The electrical component for storing energy is a capacitor. The simplest circuit for a modulator corresponds to the schematic shown in Fig. 9.

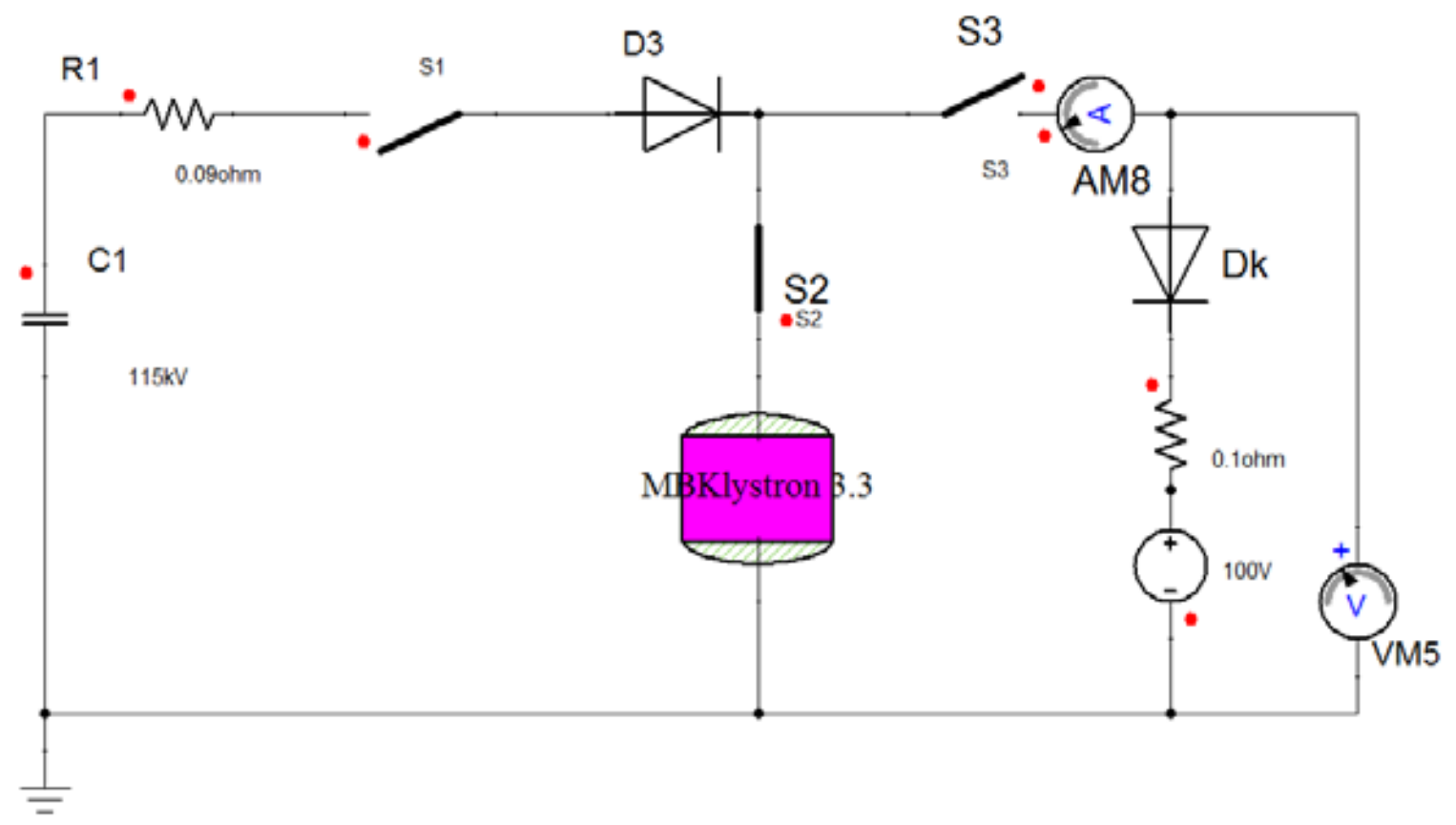

**Fig. 9:** Direct switching modulator

The voltage is 115 kV. As soon as switch S1 closes, a current flows in the klystron. The klystron at this voltage is equivalent to a resistor of 900 Ω, as shown in Fig. 4. During the pulse the voltage at the capacitor decreases, with an exponential decay of the RC discharge curve.

The advantage of a direct-switching modulator is the small number of components; however, the HV capacitor bank and the HV switch are problematic. There are very few suppliers of such a switch. Additionally, the stored energy in the system is very high, as shown in the following calculation.

Assuming the design values of the XFEL pulse (1.7 ms) with a flat top of 0.5%,

$$U = U_0 \times \mathrm{e}^{-\frac{t}{RC}} \tag{4}$$

where $U_0$ = 115 kV, $R$ = 900 Ω and $t$ = 1.7 ms,

$$\Delta U = 0.5\% = 0.005 \tag{5}$$

$$0.995 = \mathrm{e}^{-\frac{t}{RC}} \tag{6}$$

$$\ln(0.995) = -t/RC \tag{7}$$

$$C = -t/(R \times \ln(0.995)) \tag{8}$$

$$C = 377\ \mu F\ . \tag{9}$$

For the stored energy of the capacitor,

$$E_{\text{stored}} = \frac{1}{2} \times C \times U^2 \tag{10}$$

$$E_{\text{stored}} = \frac{1}{2} \times 377\ \mu\text{F} \times 115\ \text{kV}^2 \tag{11}$$

$$E_{\text{stored}} = 2491.8\ \text{kJ}\ . \tag{12}$$

To calculate the pulse energy, the pulse is simplified to a rectangular waveform,

$$E_{\text{pulse}} = U \times I \times t = \frac{U^2}{R} \times t \tag{13}$$

$$E_{\text{pulse}} = \frac{115\ \text{kV}^2}{900\ \Omega} \times 1.7\ \text{ms} \tag{14}$$

$$E_{\text{pulse}} = 24.98\ \text{kJ}\ . \tag{15}$$

The ratio between $E_{\text{pulse}}/E_{\text{stored}}$ is a factor of 100. It is obvious that this solution will be much too expensive for a series of modulators.

With reduced requirements this modulator type was realized for the ISIS front-end test stand [3]. The direct-switching modulator was built by Diversified Technologies, Inc, USA (Fig. 10).

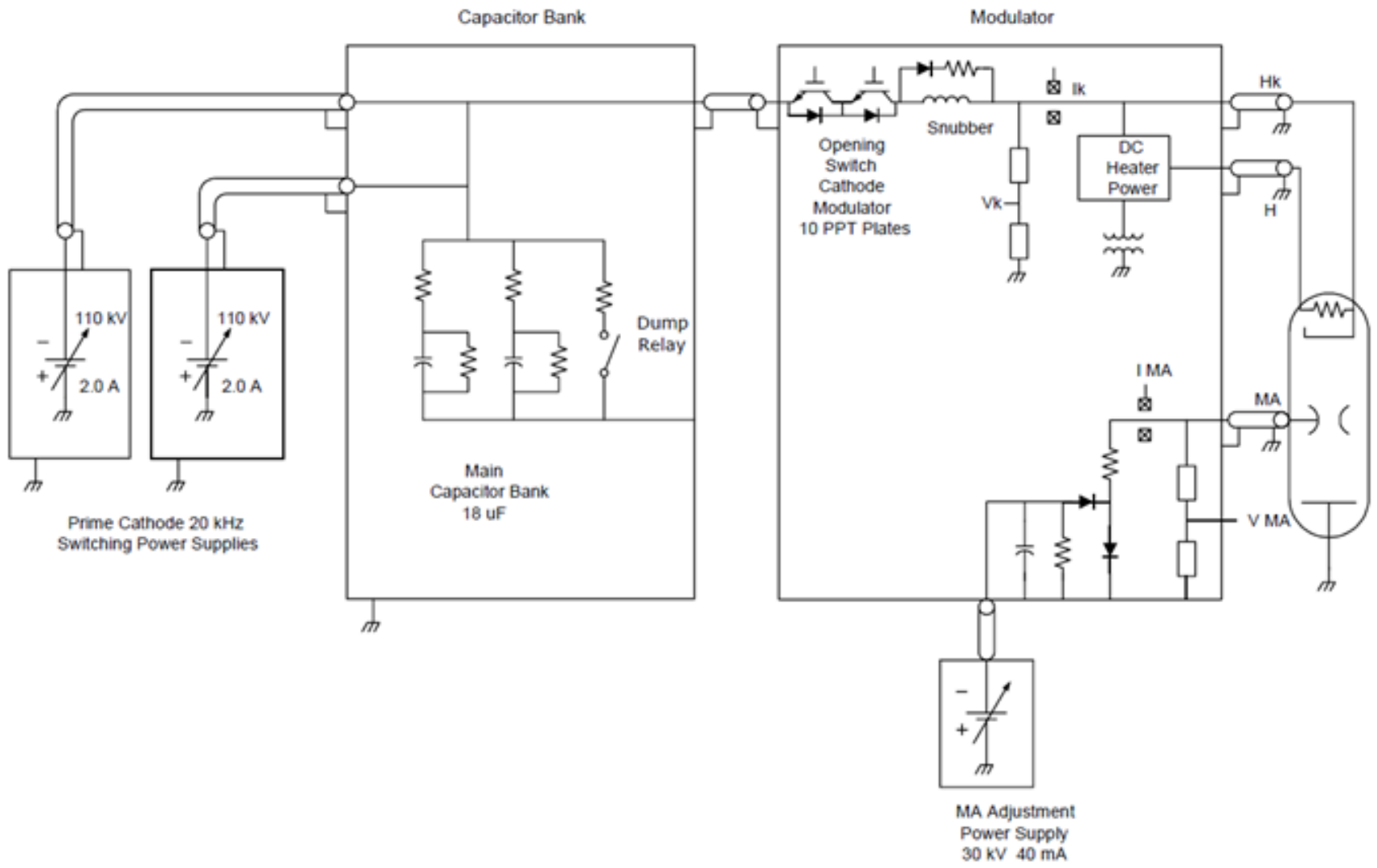

**Fig. 10:** ISIS test stand modulator [3]

### 4.1.1 ISIS modulator specifications

Table 3 contains the specifications for the ISIS modulator.

**Table 3:** ISIS modulator specifications

| Parameter | Value |
|---|---|
| Cathode voltage | −110 kV |
| Cathode current | 45 A |
| PRF | 50 Hz |
| Beam pulse width | 500 μs to 2.0 ms |
| Droop | 5% |

## 4.2 Modulator with pulse transformer

Since the design of a high-voltage switch is demanding, pulse transformer solutions were developed. A schematic is shown in Fig. 11. The primary voltage may be chosen freely, since the transformer will adapt to the required high voltage. For XFEL, 10 kV was chosen. Medium-voltage switches can be built more easily by stacking semiconductors. The first FNAL modulator used gate turn-off thyristors (GTOs), the next ones changed to insulated gate bi-polar transistors (IGBTs), and currently integrated gate-commutated thyristors (IGCTs) are used. Components that are able to handle voltages up to 6.5 kV are currently on the market. The semiconductor switch can be built without oil, but the pulse transformer has still to be under oil.

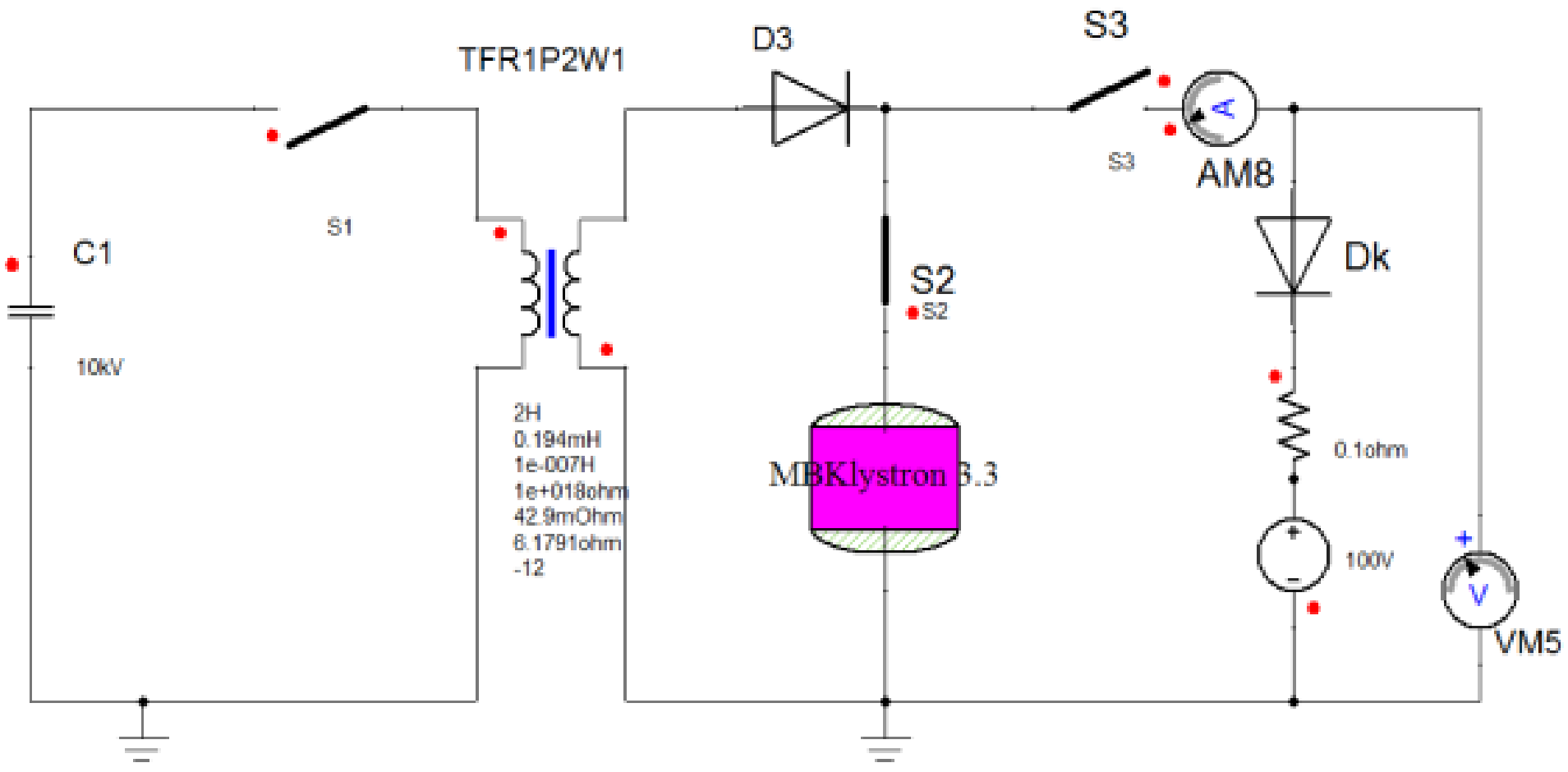

**Fig. 11:** Modulator with pulse transformer

The disadvantages are the volume and weight of the transformer. The additional leakage inductances increase the rise time, and even more energy is stored in the system, which has to be dissipated in the event of an arc.

According to the equivalent circuit for a transformer (Fig. 12) several inductances can store energy during the pulse. The stray inductances carry the main pulse current and short-circuit current of up to 2000 A, whereby the main inductance is loaded by the applied voltage of 10 kV. The stored energy is maximal at the end of the pulse.

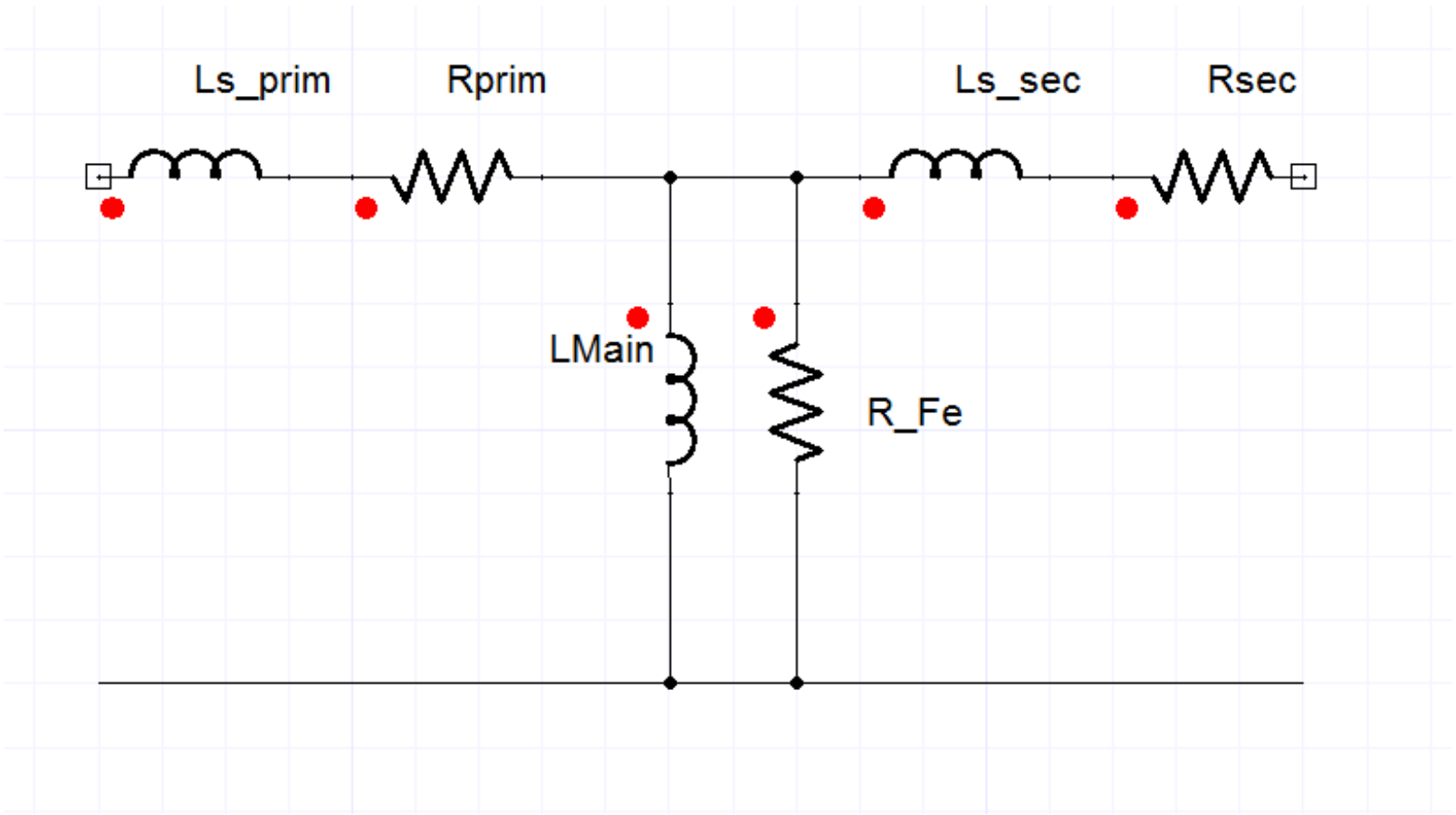

**Fig. 12:** Equivalent circuit for a transformer

Stored energy in the stray inductances are given by

$$E_{\text{stored Ls}} = \frac{1}{2} \times Ls \times I_{\text{short circuit}}{}^{2} , \tag{16}$$

where $L_s$ for the XFEL transformer = 200 µH,

$$E_{\text{stored Ls}} = \frac{1}{2} \times 200\ \mu\text{H} \times 2000\ \text{A}^{2} , \tag{17}$$

$$E_{\text{stored Ls}} = 400\, J , \tag{18}$$

and the stored energy in the main inductance is given by

$$E_{\text{stored LM}} = \frac{1}{2} \times L \times I_{\text{main}}^{2} \tag{19}$$

$L_{\text{Main}}$ of the XFEL transformer = 5 H,

$$I_{\text{Main}} = \frac{U \times t}{L}\,, \tag{20}$$

and $U$ = 10 kV, $t$ = time of arc 0–1.7ms,

$$I_{\text{Mainmax}} = \frac{10\ \text{kV} \times 1.7\ \text{ms}}{5\ \text{H}} = 3.4\ \text{A}\,, \tag{21}$$

$$E_{\text{stored LM}} = \frac{1}{2} \times 5\ \text{H} \times 3.4\ \text{A}^{2} = 28.9\ \text{J}\,. \tag{22}$$

In total, 428.9 J are stored in the transformer. The energy deposited in the klystron has to stay below 20 J. Therefore, a pulse transformer demands the introduction of the possibility of discharging. A simple circuit such as an RCD network as shown in Fig. 13 comprising Rvar, R8, C3 and a diode. For energy dissipation the resistance of the secondary winding also has to be taken into account.

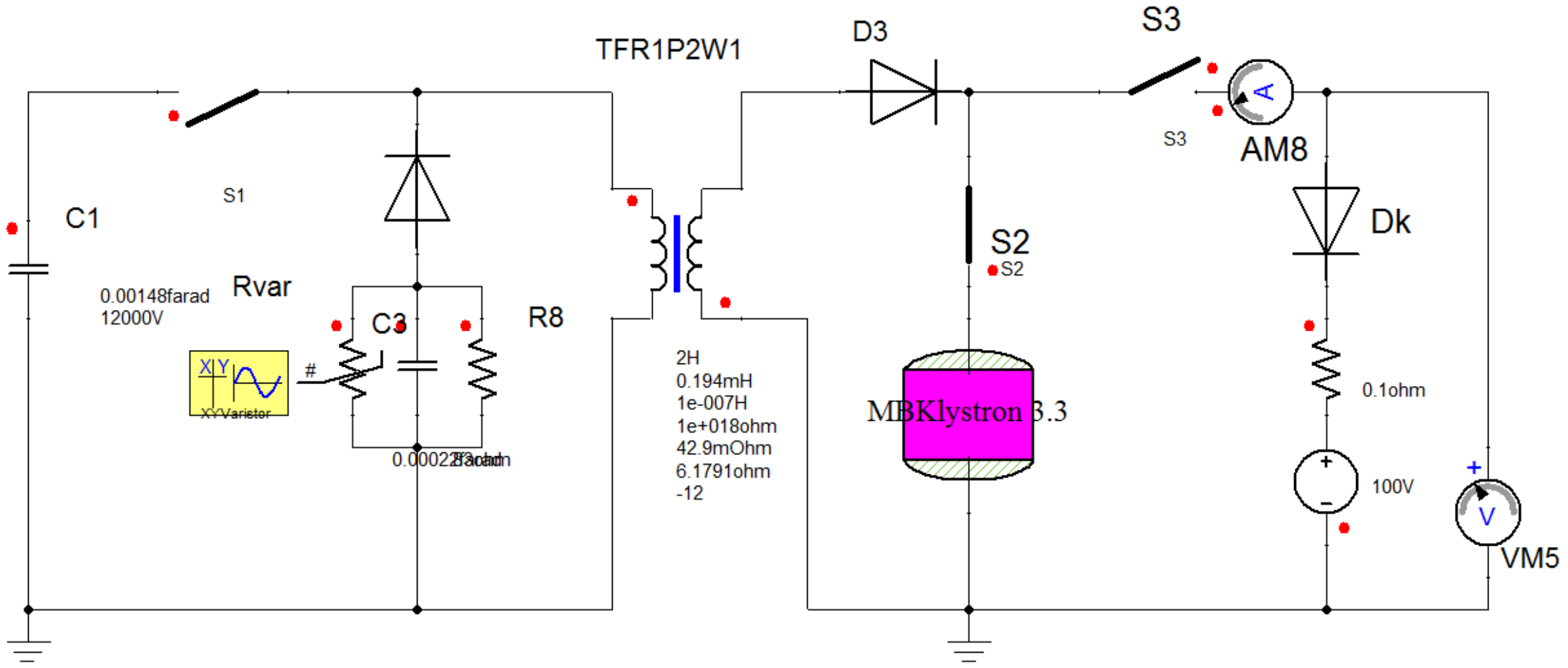

**Fig. 13:** Modulator with pulse transformer with discharge network

Note that the introduction of the pulse transformer does not affect the calculation of the stored energy inside the modulator. Only the voltage levels at the capacitor bank and the semiconductor have changed.

## 4.3 Bouncer modulator

To decrease stored energy, the bouncer modulator was developed at Fermi Lab [1]. An LC ringing circuit was introduced into the circuit as shown in Fig. 14. It compensates for the voltage droop at the capacitor bank during the pulse; Kirchhoff's law explains this behaviour. The sum of all voltages in a circuit equals zero. The voltage at the input of the pulse transformer plus bouncer voltage is equal to the capacitor voltage. If capacitor voltage and bouncer voltage now decrease by the same amount, the transformer voltage stays constant, as shown in Figs. 15 and 16.

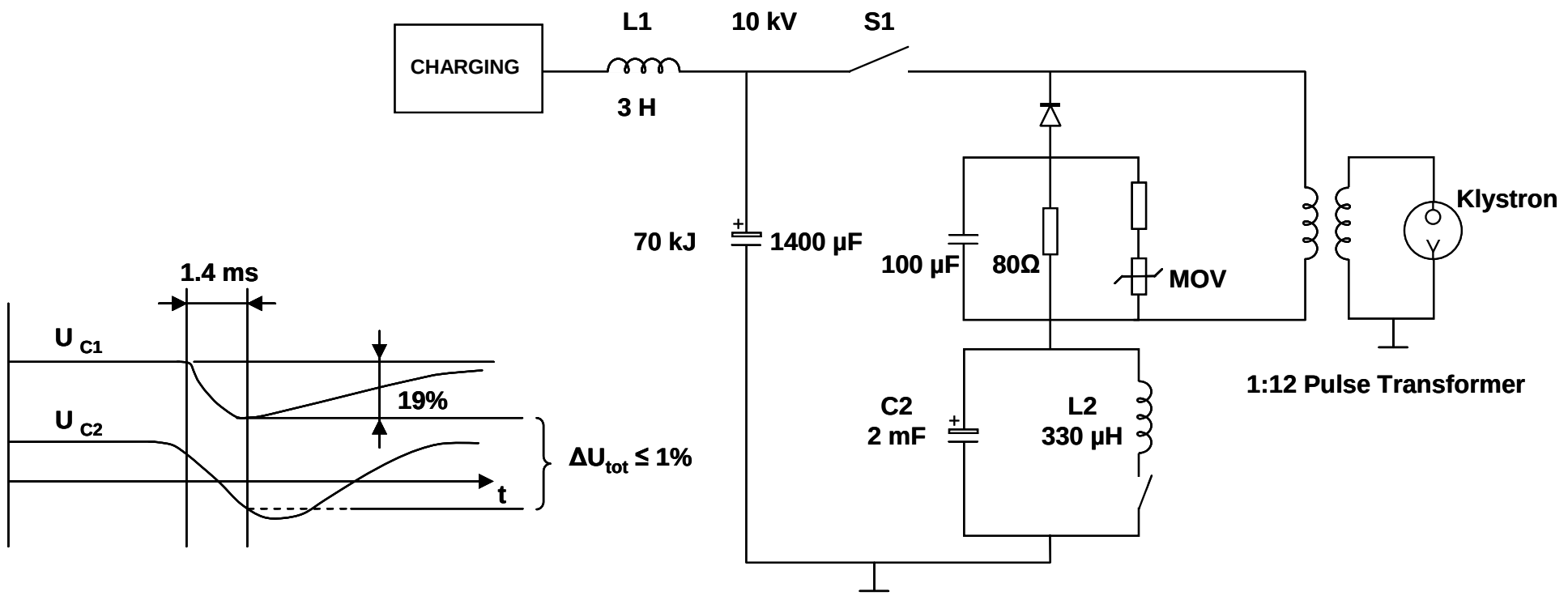

**Fig. 14:** Bouncer modulator

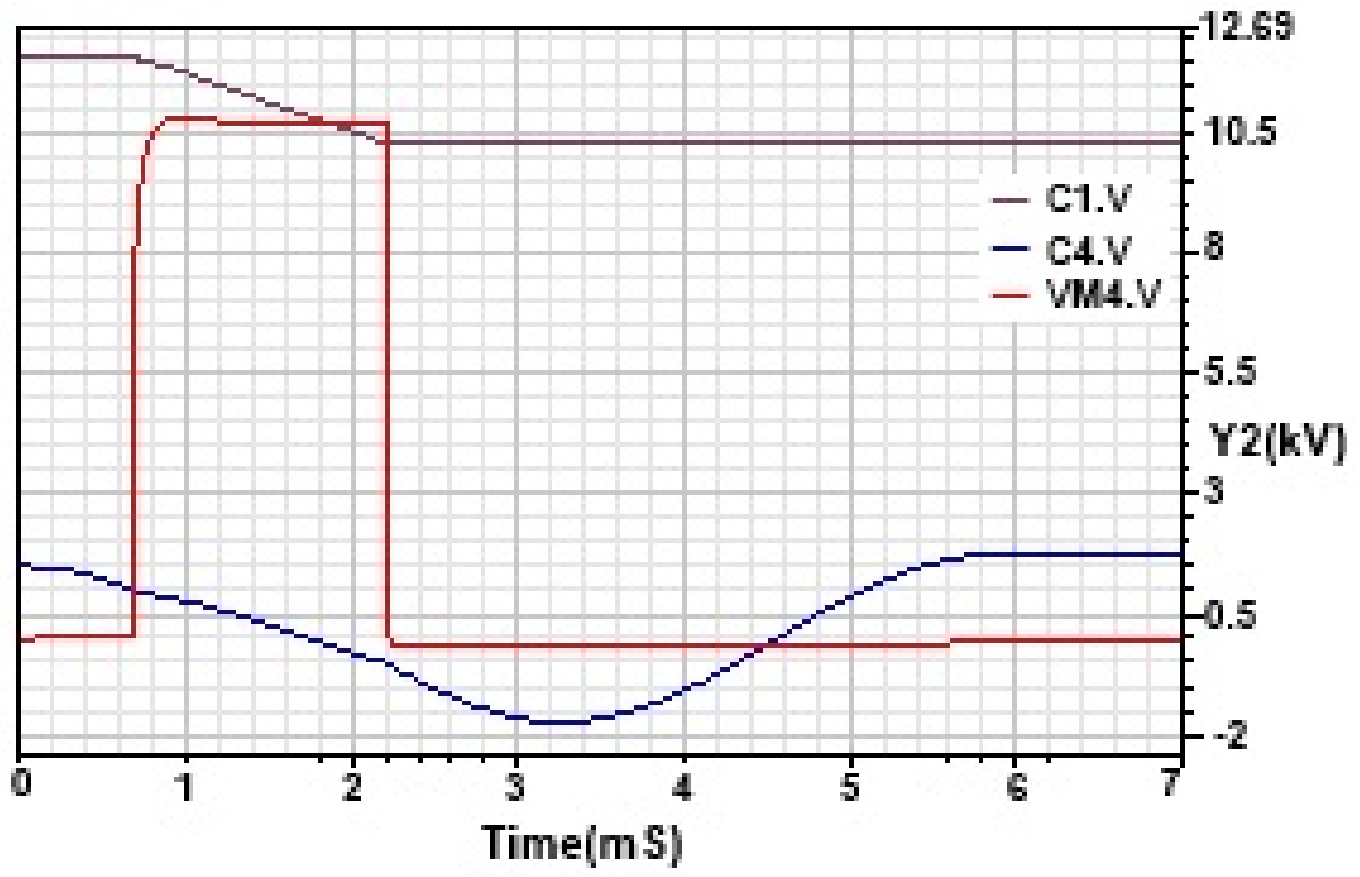

**Fig. 15:** Voltage waveforms in a bouncer modulator. Red: voltage at the transformer; violet, capacitor bank voltage; blue, bouncer cap voltage.

Since the capacitor voltage droop is an exponential decay and the bouncer oscillates with a cosine waveform, the voltages do not droop completely equally, leaving a ripple on the flat top as shown in Fig. 16.

This ripple is adjustable by the dimensioning of the bouncer elements, the timing between the main switch and bouncer switch and the voltage of the bouncer.

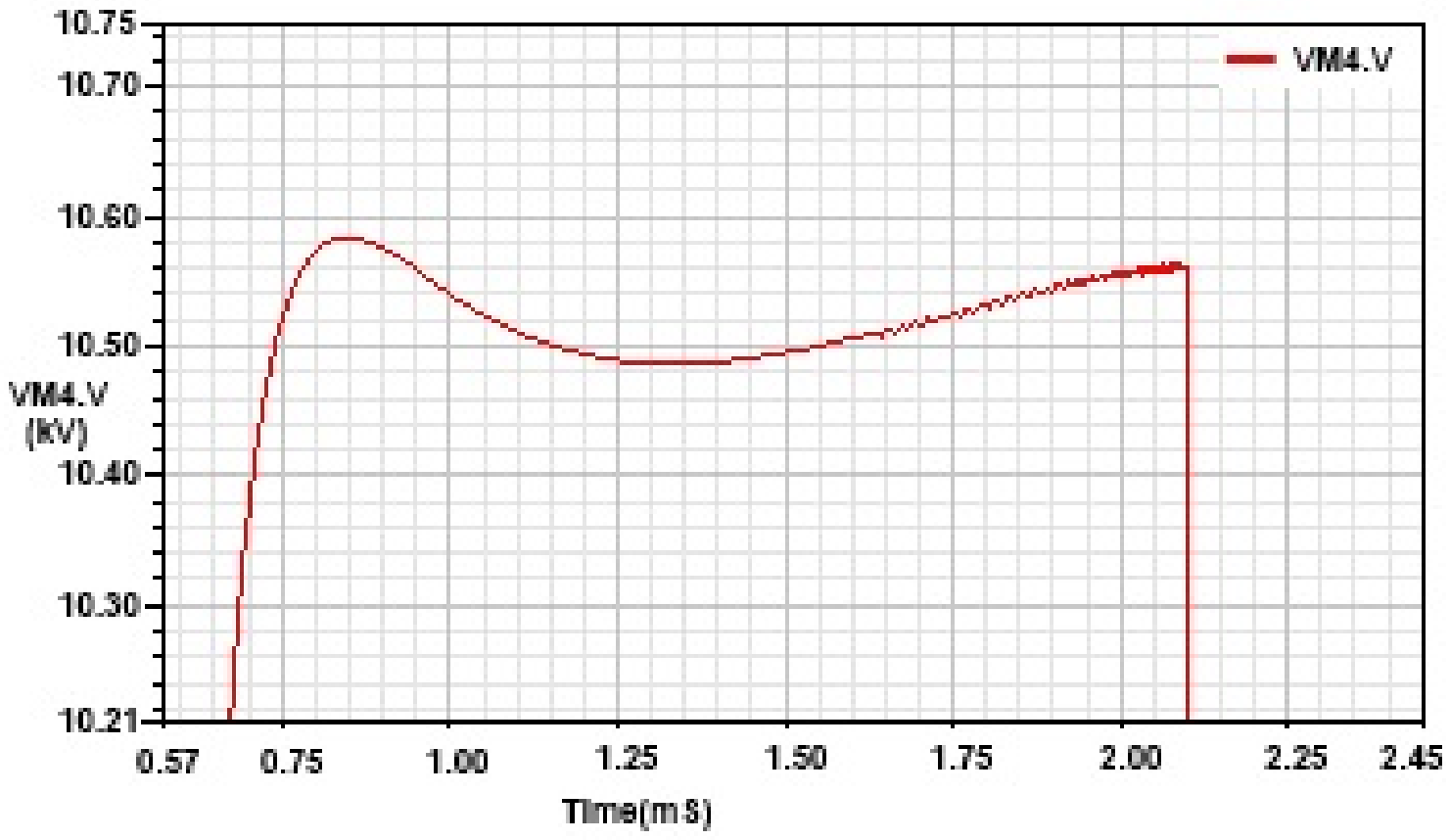

**Fig. 16:** Voltage ripple at the flat top of a bouncer modulator

To see the advantage of the bouncer principle, a calculation of the stored energy is given below. The stored energy in the modulator is:

Main capacitor energy:

$$E_{\text{stored}} = \frac{1}{2} \times C_{\text{main}} \times U^2 \,, \tag{23}$$

$$E_{\text{stored}} = \frac{1}{2} \times 1.4 \text{ mF} \times 10 \text{ kV}^2 \,, \tag{24}$$

$$E_{\text{stored}} = 70 \text{ kJ} \,. \tag{25}$$

Bouncer energy:

$$E_{\text{bouncer}} = \frac{1}{2} \times C_{\text{bouncer}} \times U_{\text{bouncer}}{}^2 \,, \tag{26}$$

$$E_{\text{bouncer}} = \frac{1}{2} \times 2 \text{ mF} \times 2 \text{ kV}^2 \,, \tag{27}$$

$$E_{\text{bouncer}} = 4 \text{ kJ} \,. \tag{28}$$

The sum is 74 kJ, which is only five times the pulse energy.

The first three bouncer modulators were built by FNAL and operated at DESY. At DESY further research and development was done and the modulator was brought to industry. The following modulators were built by Puls-Plasmatechnik GmbH (PPT), Germany, the nowadays Ampegon PPT GmbH [4] and are operated with the FLASH accelerator and test stands for the XFEL. Improvements were made for reducing EMI basically by improving the construction e.g. using sandwich structures for the current leads. A constant power charging of the capacitor bank as described in section 6 was introduced. The bouncer changed position to the high-voltage side. Because of this the pulse transformer primary stays at ground potential. In accordance with Kirchhoff's law the position of the different voltages does not influence the transformer voltage.

A disadvantage of the bouncer is the high current cycling in the LC network. Additionally, a new time-dependence of the triggering of the switch is introduced.

The bouncer principle was later also proposed by other enterprises such as, for example, JEMA, Spain [5] (Fig. 17). Here the single-pulse transformer is split into several smaller primary windings and one secondary winding. This approach takes advantage of the low energy storage due to the bouncer principle. Due to the splitting of the primaries the semiconductor switches do not have to handle the high pulse current. Also, the lower stray inductances in each branch result in a better rise time of the pulse.

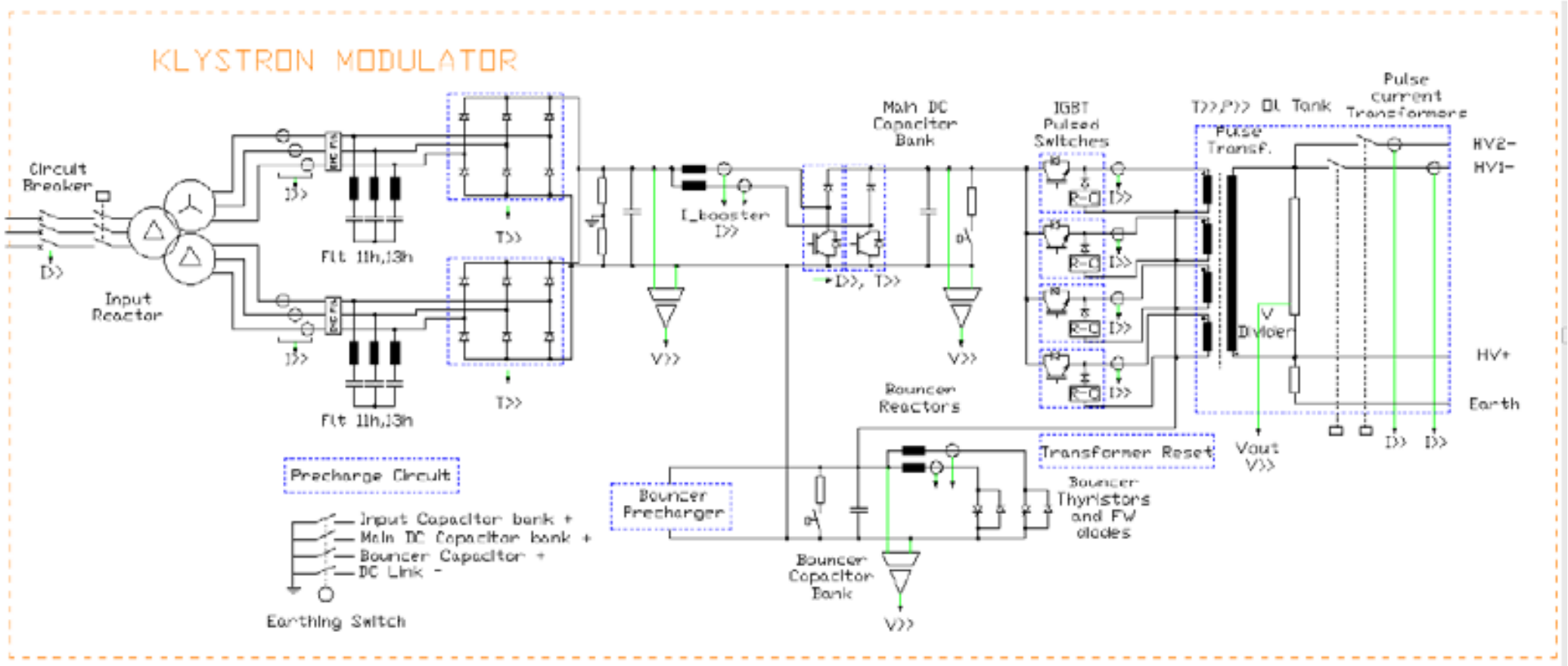

**Fig. 17:** Proposal for a klystron modulator by JEMA, using the bouncer principle

### 4.4 Active bouncer solutions

To further decrease the remaining ringing of the flat top using the LC bouncer, switched mode power supplies can be introduced. Here, high-frequency switching semiconductors have to be used. The development shown in Fig. 18 was done at CERN [6]. This topology was tested as a prototype and tests are ongoing. Other topologies for switched mode supplies are also possible.

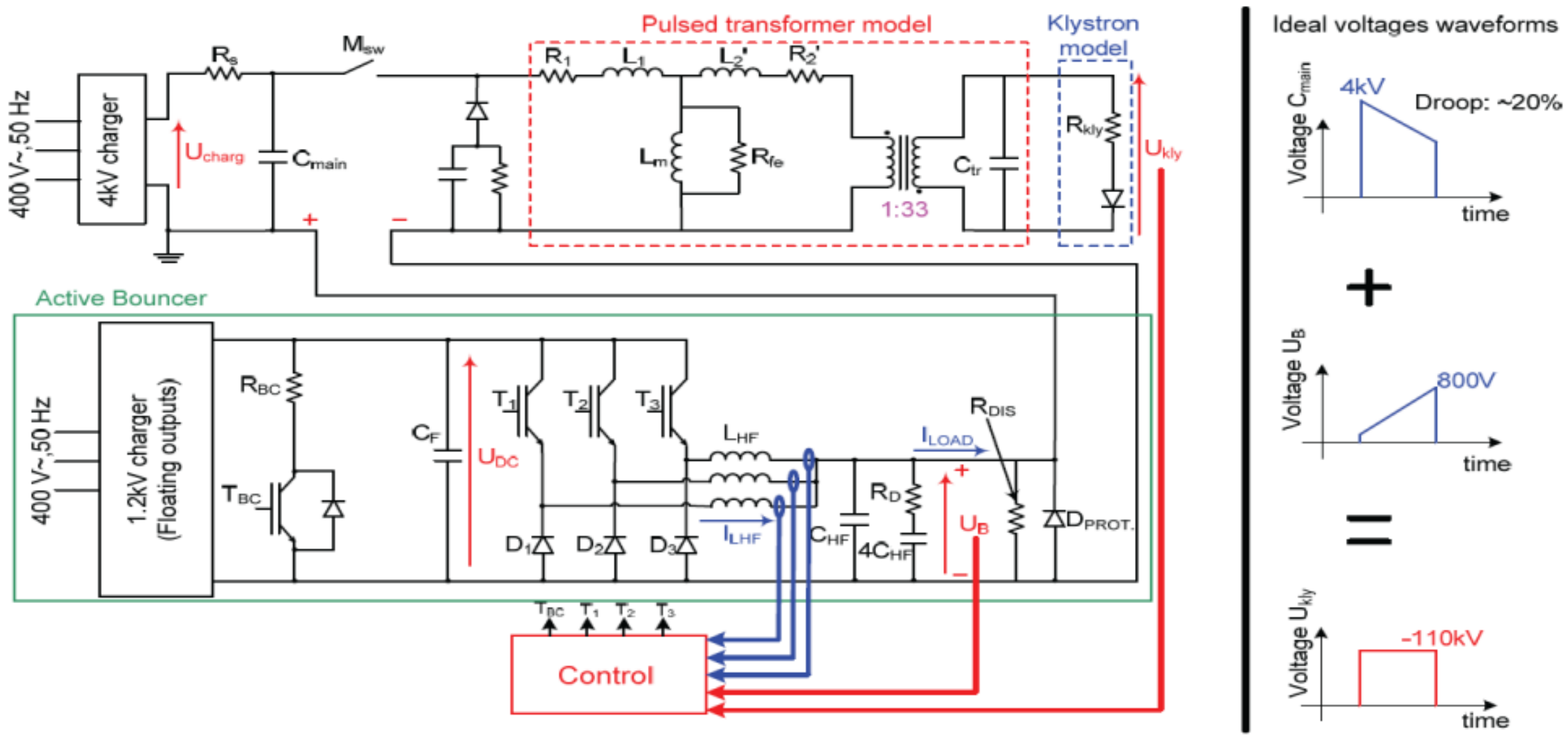

**Fig. 18:** Active bouncer solution tested at CERN [6]

### 4.5 Pulse forming by LC circuit

Another passive pulse-forming topology, instead of a LC bouncer, is an LR circuit, as shown in Fig. 19. Here time-dependent current flow in the inductance during the pulse is used. At the beginning of the pulse, the inductance has a high impedance. The current flows through the resistor having an initial voltage drop. During the pulse the current commutates into the inductance, which short-circuits the resistor. This solution is built by Scandinova, Sweden and PPT, Germany. In Fig. 20 the schematic for a Scandinova modulator is presented.

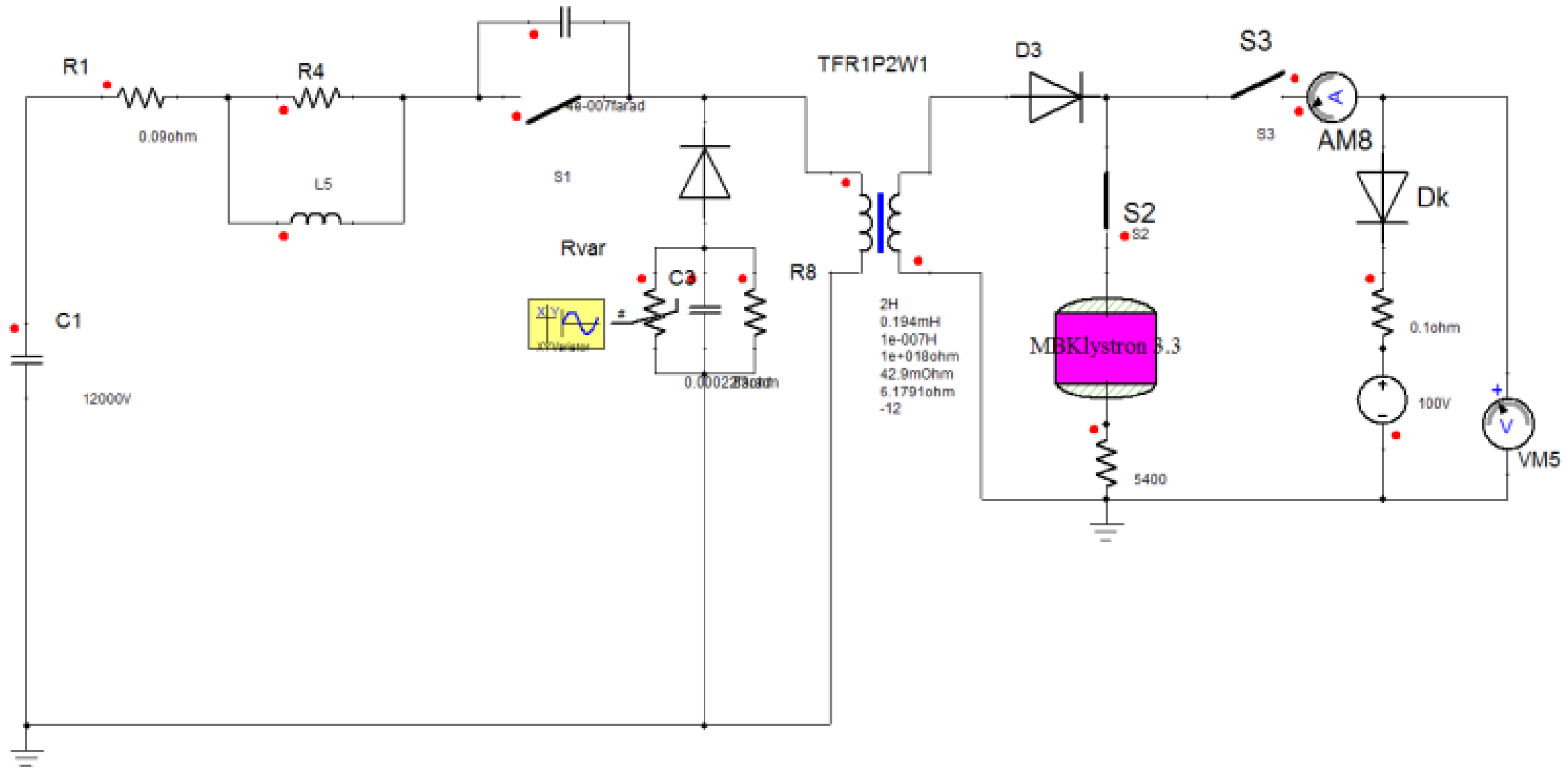

**Fig. 19:** Pulse forming by LR network

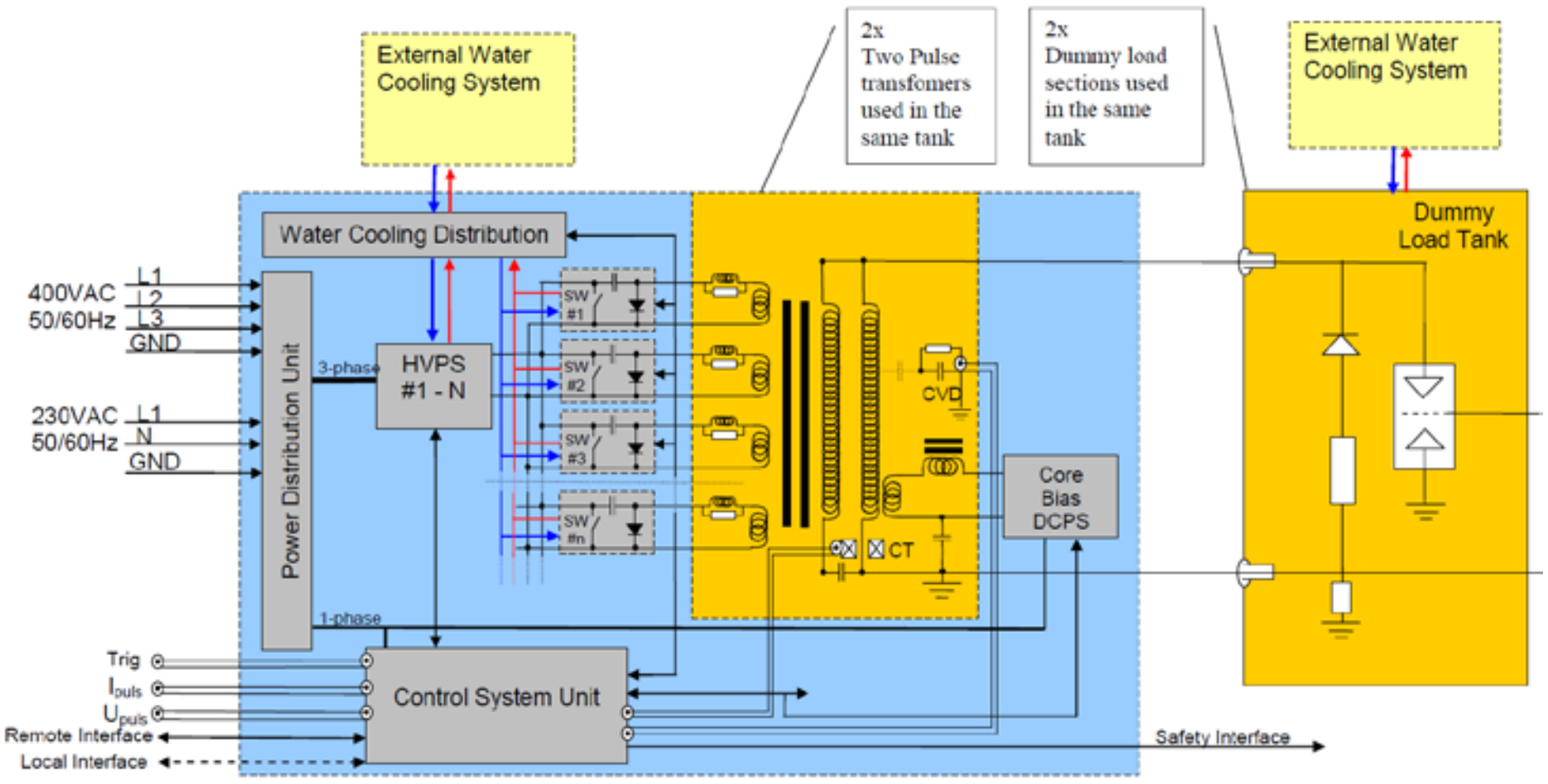

**Fig. 20:** Modulator with LR pulse-forming network

The advantage of the LR is definitely its simplicity. Once the LR circuit is optimized it does not require any further timing signals. However, it is also less flexible than the bouncer, where over a small range some changes in pulse shaping are possible. The additional losses in the resistor have to be taken into account for the efficiency calculation.

## 5 Modulators with switched mode power supplies

The development of better semiconductors and transformers has led to different solutions for modulators using switched mode power supplies for pulse generation. In this section three solutions are presented. These modulators are either already installed or prototypes have been proven to work.

### 5.1 Pulse step modulator

This technology was developed by Ampegon, Turgi, Switzerland [7, 8]. The original application was for shortwave transmitters to produce amplitude-modulated signals. A rectangular pulse is 'just' a special waveform that can be produced with this topology.

In the modulators, several voltage sources are stacked in series, as shown in Fig. 21. By means of appropriate regulation and time pattern the voltage sources are turned on to produce the required waveform.

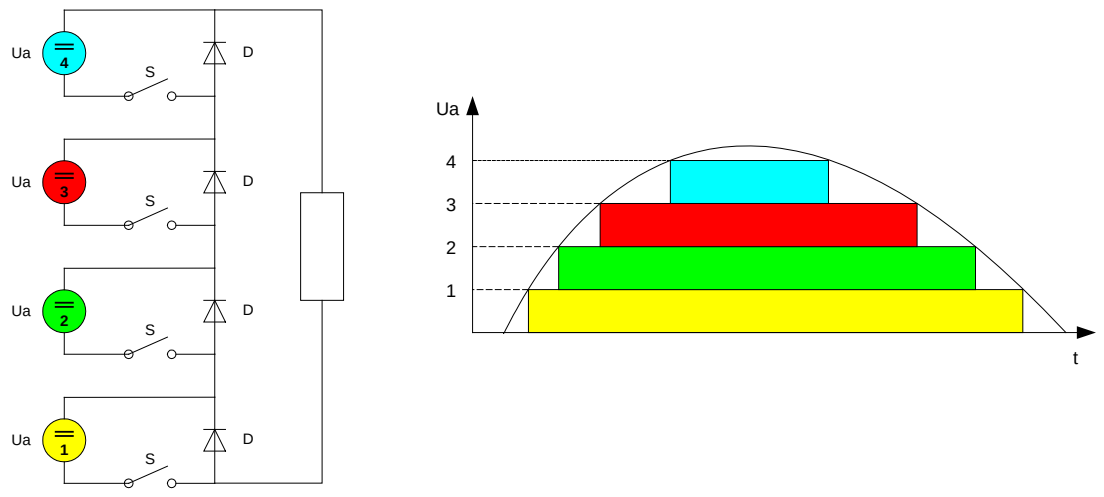

**Fig. 21:** Principle of the pulse step modulator

As shown in Fig. 21, the simple turning on and off of the voltage source would lead to a large quantification error. This is well known from digital signal processing. To achieve a more precise adaptation of the desired wave form, switched mode supplies are used. The principle shown in Fig. 22 is also described in Ref. [9].

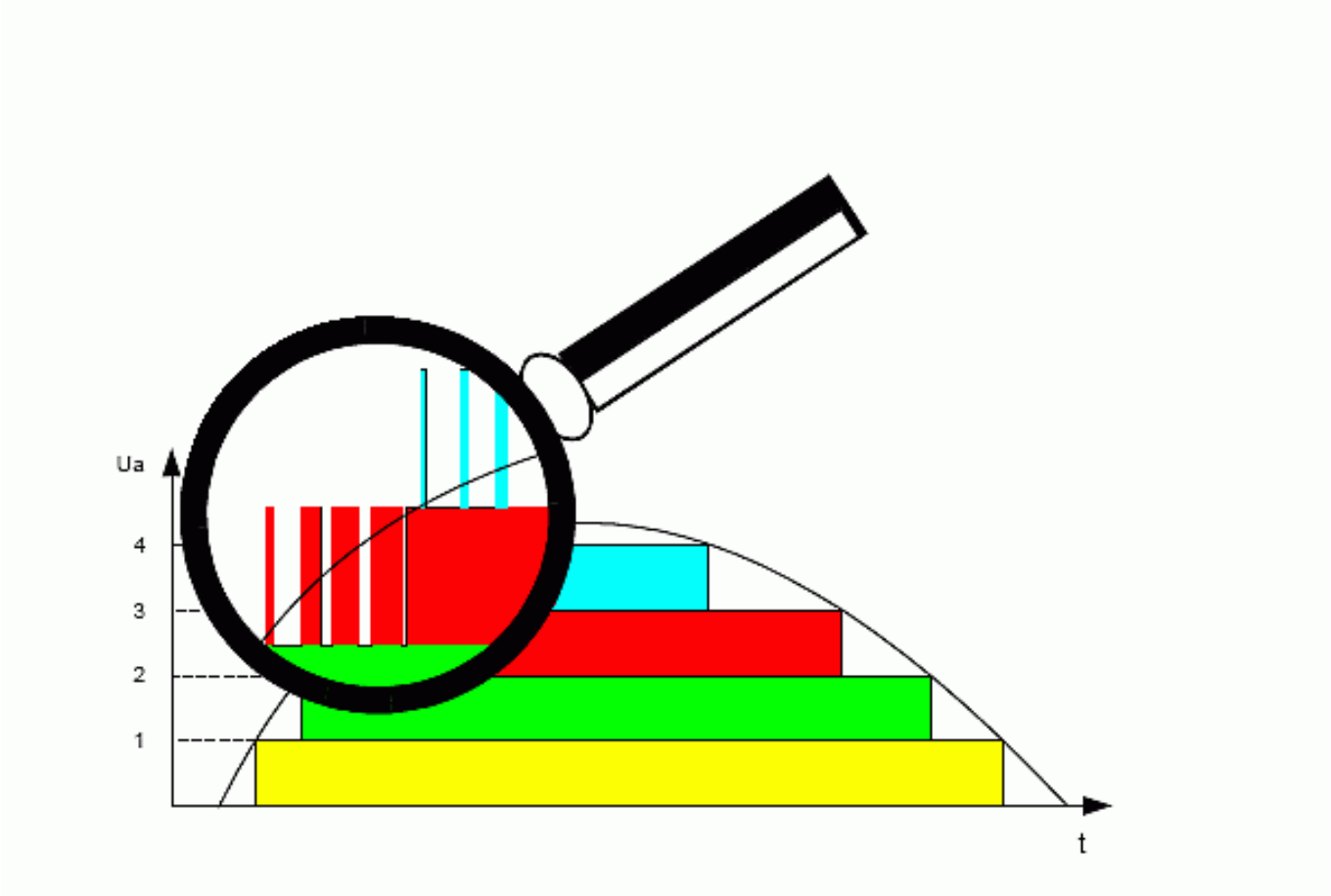

**Fig. 22:** Use of switched mode supplies to better shape the wave form

A complete RF station is presented in Fig. 23. The stacked voltage sources are built next to the switching modules. A low-pass filter is introduced to decrease the voltage ripple and suppress the switching frequency at the output of the modulator.

Specific to DESY is the use of pulse cables to transport the pulses from the modulator to the pulse transformer and klystron. The length of these cables is up to 1.7 km [10].

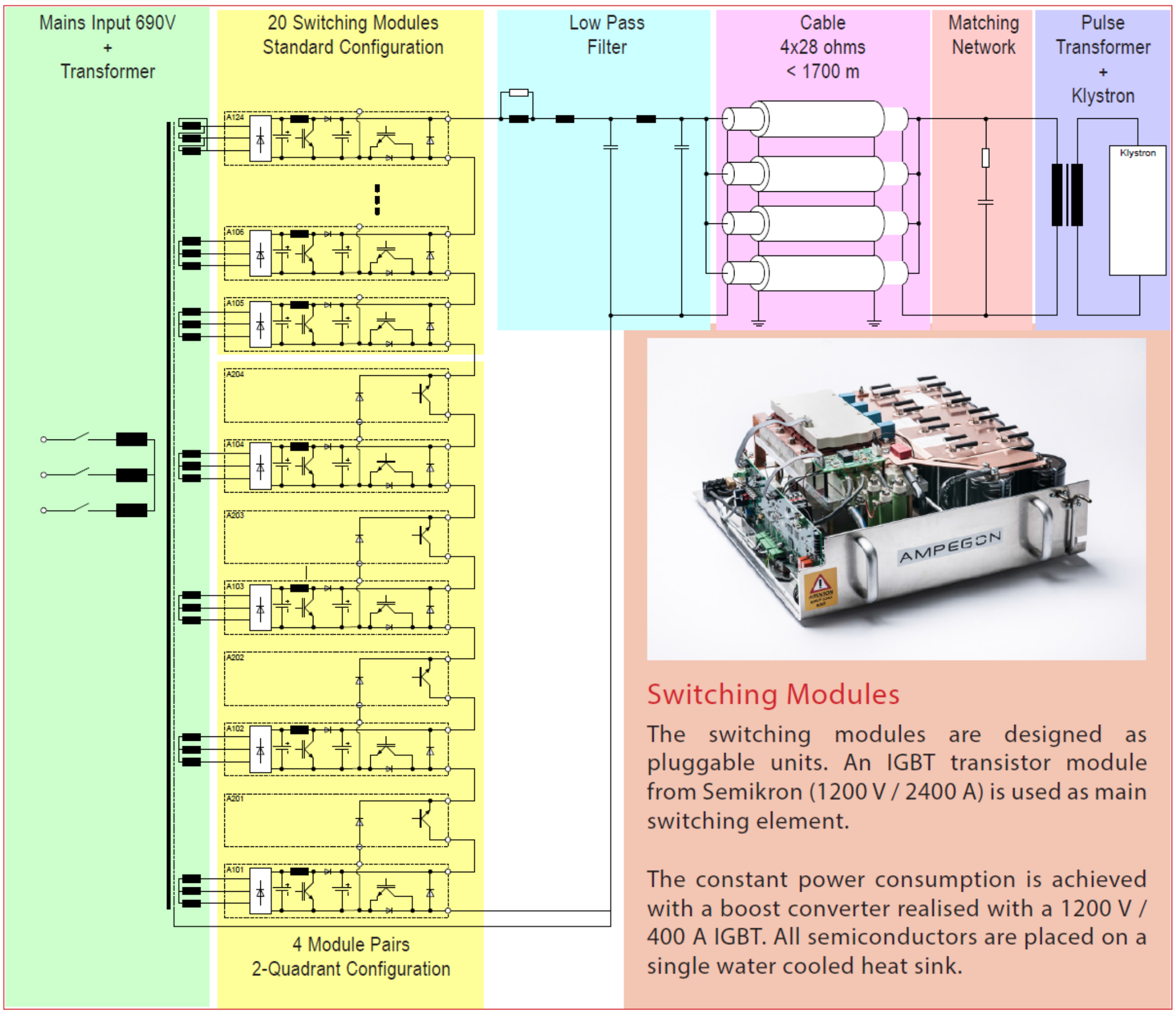

**Fig. 23:** Schematic of the RF station with Ampegon modulator [8]

The waveform of the klystron current and voltage, as well as the modulator current and voltage are shown in Fig. 24. The rise time is approximately 200 µs.

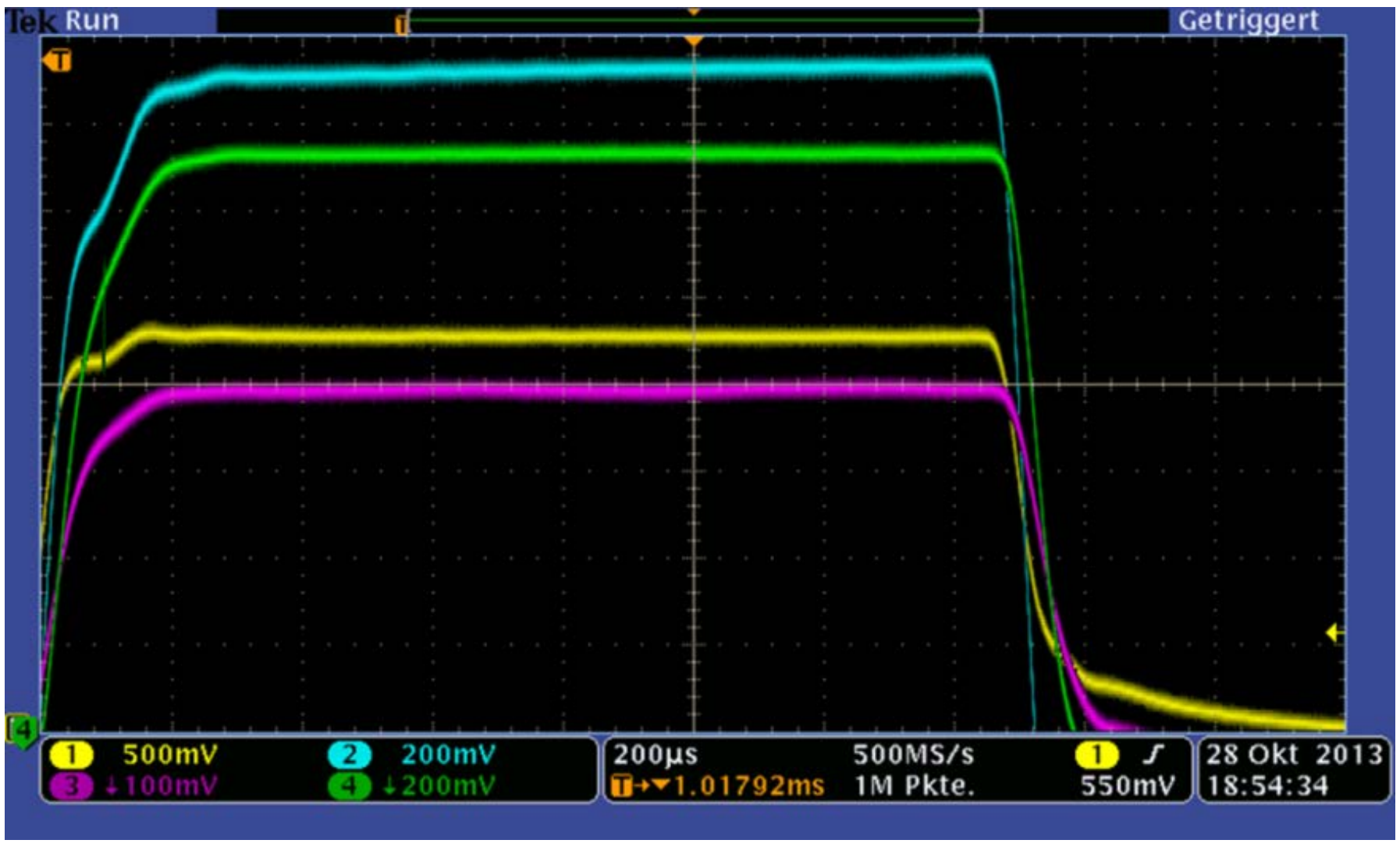

**Fig. 24:** Current waveforms of the Ampegon modulator. Yellow: modulator voltage (9kV), light blue: modulator current (1500A), pink: Klystron voltage (108 kV), green: Klystron current (125A).

The flat top is measured to be 30 $V_{pp}$ on a 10 kV pulse. This corresponds to 0.3% pp, which is a very good result and stays within the specification. The measurement is shown in Fig. 25.

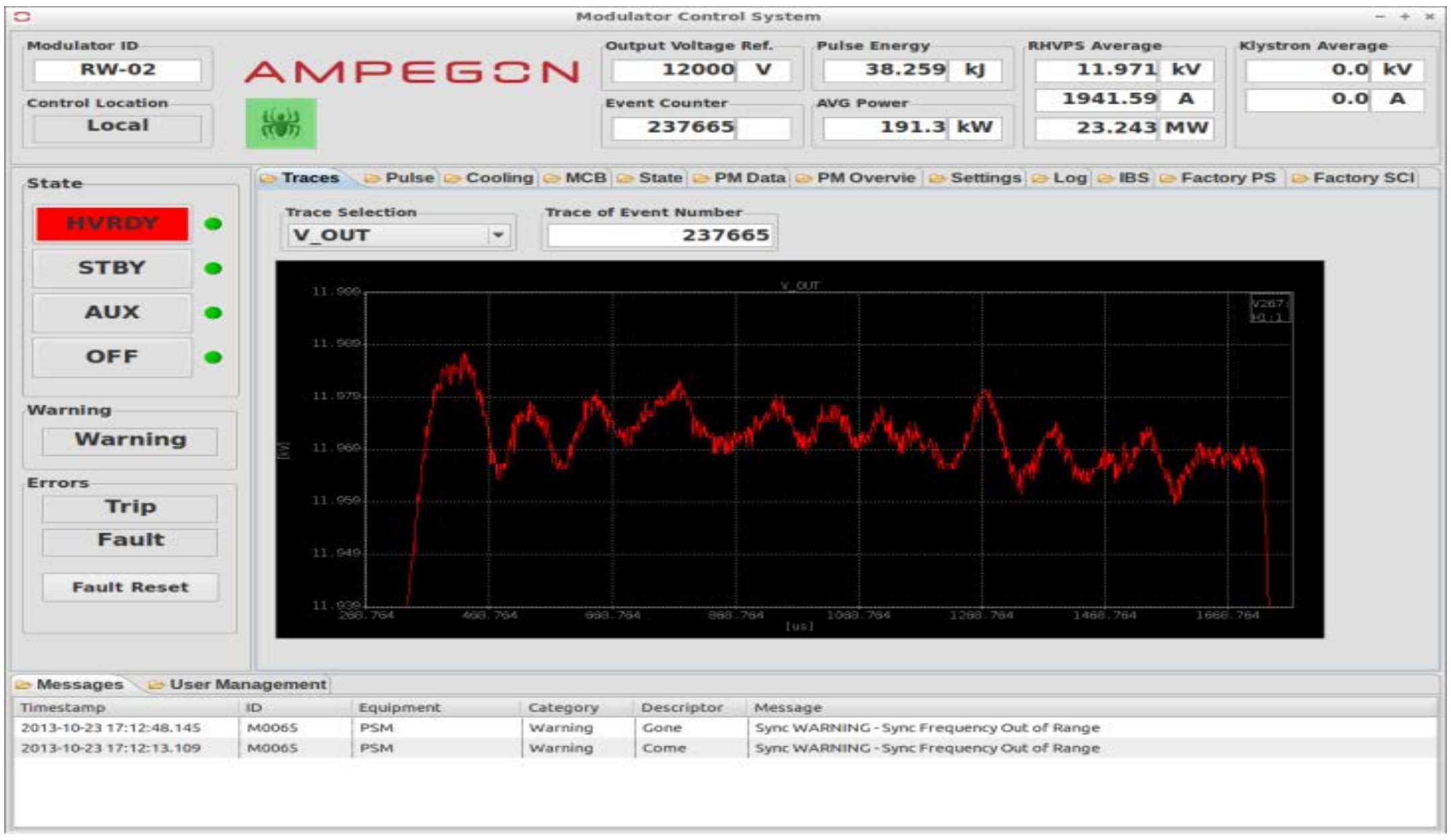

**Fig. 25:** Flat top measurement for the XFEL modulator

## 5.2 H-bridge converter/modulator

The topology chosen for the SNS modulator was H-bridge topology with interleaved switching pattern (Fig. 26); see Ref. [11]. For the input stage a three-phase SCR controller supplies a DC intermediate circuit. Via multiphase IGBT H-bridges, transformers and a secondary diode rectifier, 135 kV is generated. The switching frequency of the IGBTs is 20 kHz. The peak power is up to 11 MW with an operating repetition rate of 60 Hz at a pulse length of 1.35 ms. Figure 27 shows the modulator assembly.

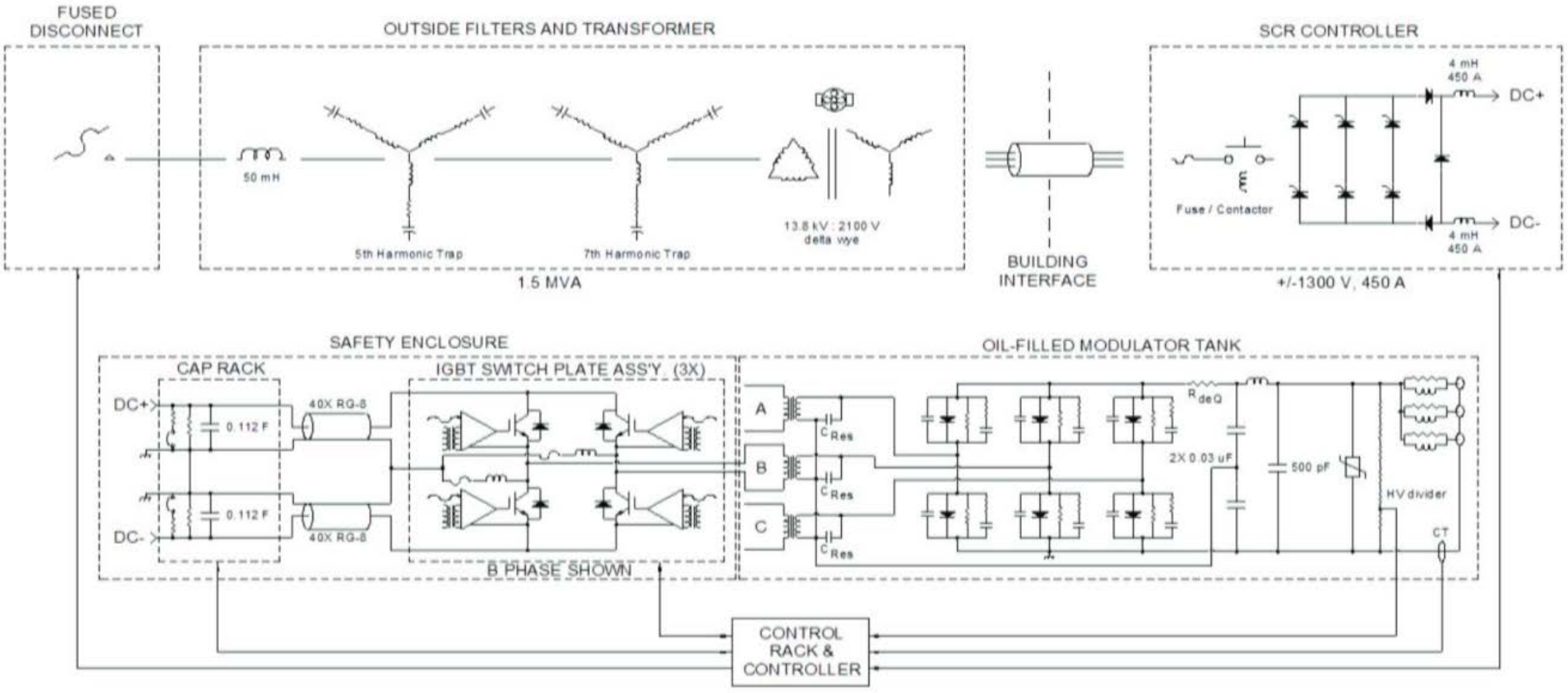

**Fig. 26:** Multiphase H-bridge SNS modulator

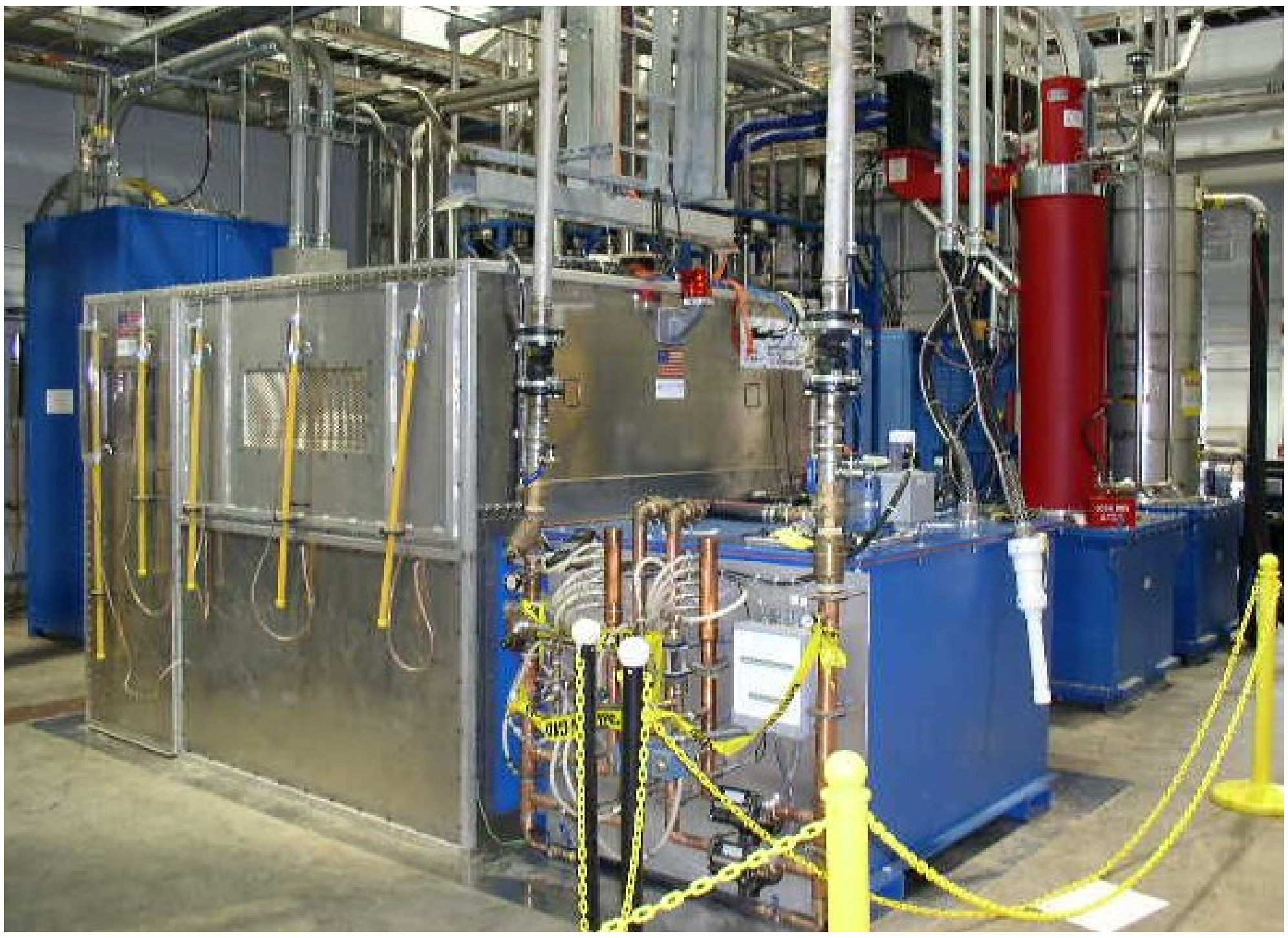

**Fig. 27:** SNS modulator assembly

### 5.3 Marx modulator

The principle of producing high voltage by using a Marx generator is well known. The parallel charging of high-voltage cells at a low voltage and then switching these in series was described in the 1920s. At SLAC this principle was used to build a modulator. The modulator is divided into modules that are loaded at low voltage [12]. During the pulse the modules are connected in series. Figure 28 presents the schematic of the modulator. An additional droop compensation circuit regulates the flat top to compensate for the storage capacitor voltage droop (Fig. 29).

This modulator was built as a prototype with full power and voltage.

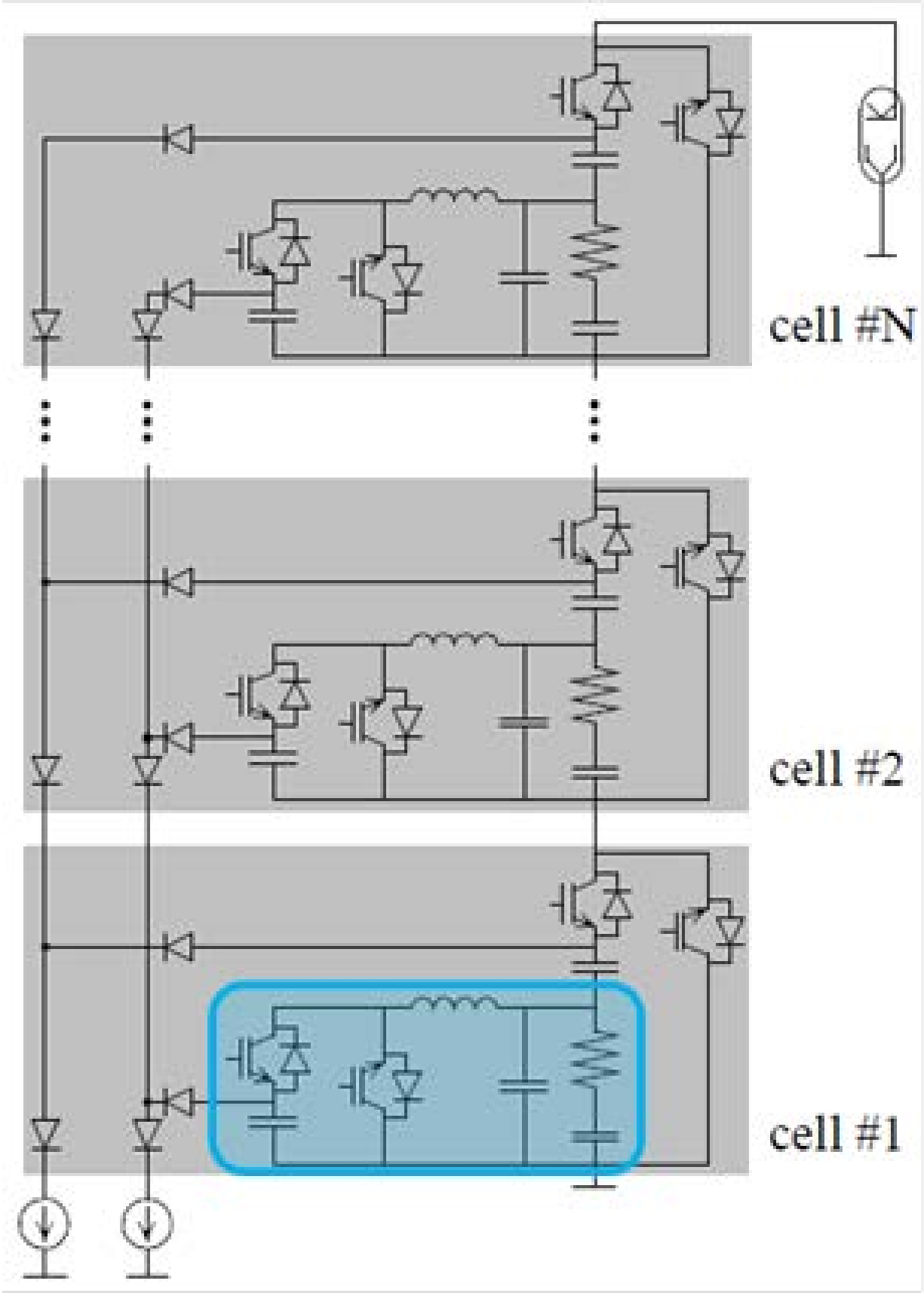

**Fig. 28:** Schematic of Marx modulator

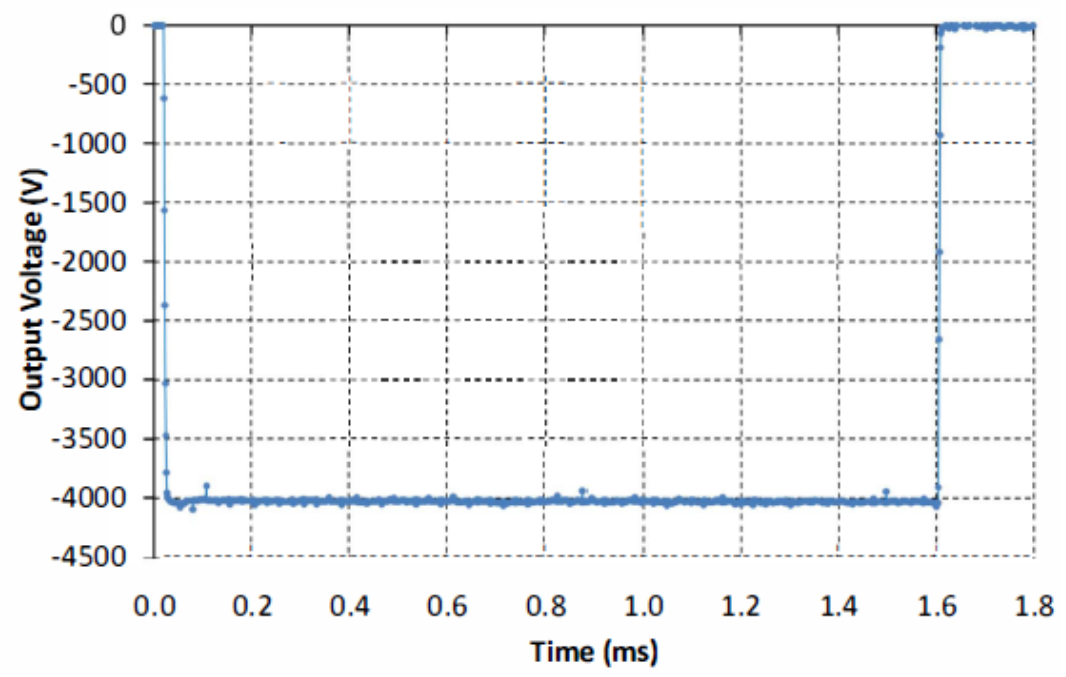

**Fig. 29:** Waveform for one cell

## 6 Connecting modulators to the electrical grid

By means of capacitors modulators constantly draw electrical energy from the mains to supply pulsed output power. As an example, the bouncer modulator is shown in Fig. 30. During the pause between pulses this energy has to be recharged. Figure 31 shows the waveform of the capacitor voltage of a bouncer modulator in pulsed operation. Recharging this circuit with either a diode rectifier or SCR rectifier would draw high peak currents, and therefore high peak power from the mains, leading to severe disturbances to the grid.

The first modulator installed at DESY used a SCR primary regulated charging unit. The pulsing of this one modulator could be measured at the ppm level in the HERA magnet power supplies. It is easy to imagine that a large number of modulators, as at XFEL with an integrated pulse power of 450 MW, cannot be connected to the grid without taking special care for compensation.

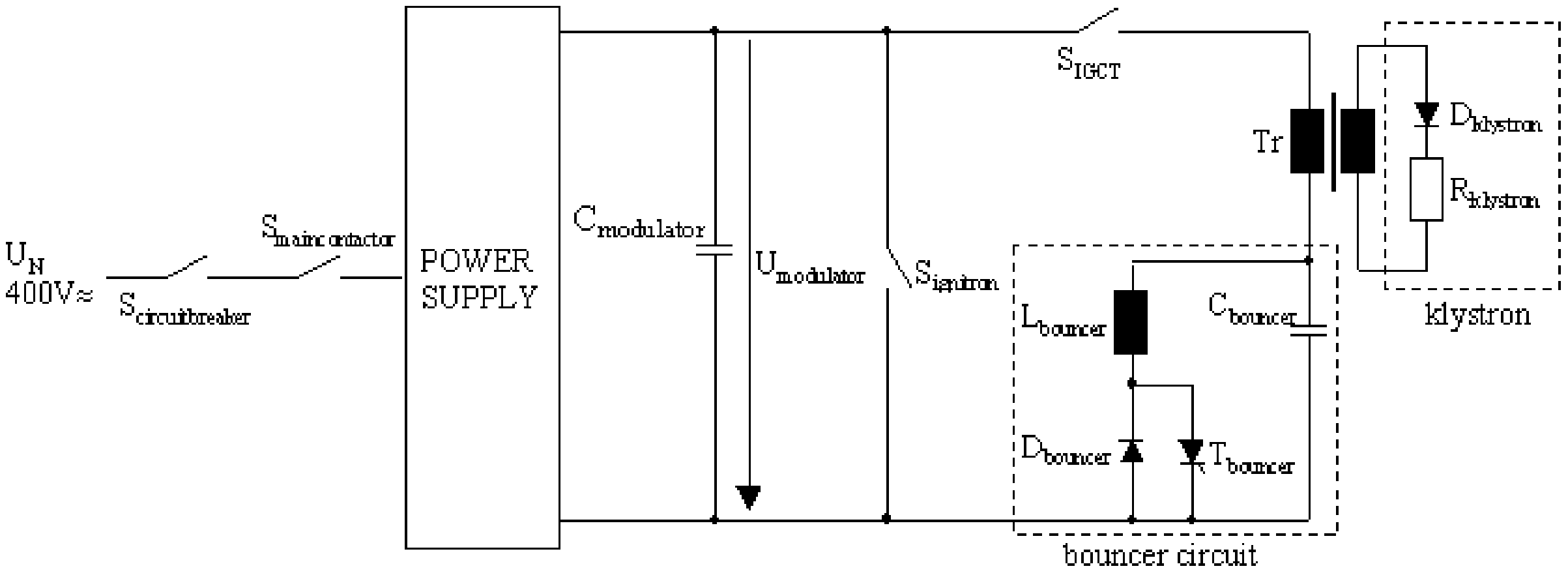

**Fig. 30:** Topology of the bouncer modulator; energy is stored in the main capacitor bank

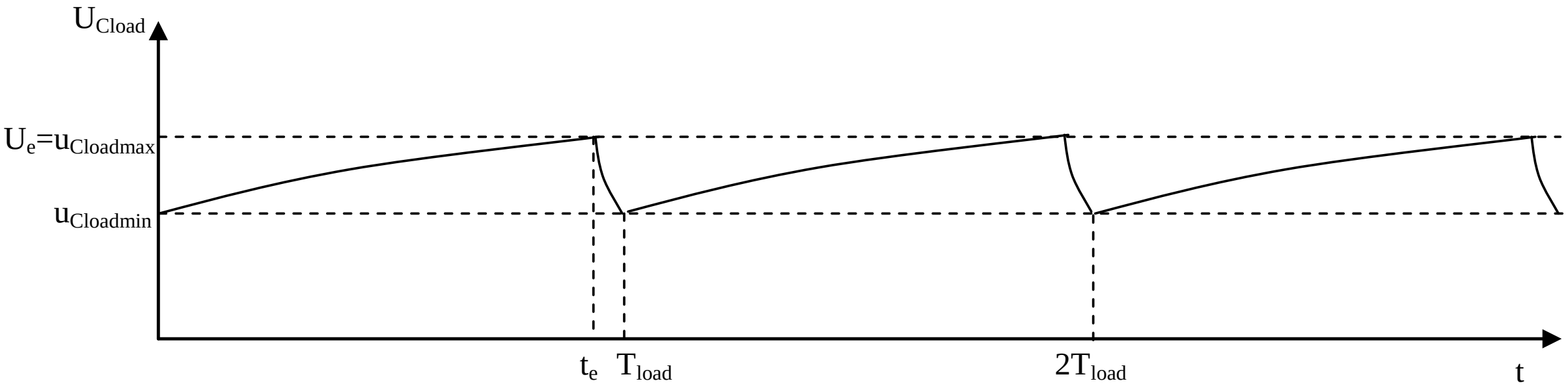

**Fig. 31:** Voltage waveform of the main capacitor bank

To prevent disturbing the consumers connected to the mains, the level of allowed disturbance is defined in the German standard VDE 0838, IEC 38 or the equivalent European standard EN 61000-3-3. In general, no consumer is allowed to produce more distortion than 3% of the voltage variation of the mains. At the low frequencies at which the distortion can be seen as a fluctuation in the intensity of lighting, these disturbances are even more restricted. At these frequencies, the human eye is very sensitive to changes in light intensities. The eye sees a flickering, which gives the name ‘flicker frequencies’. Figure 32 shows the maximum grid distortion as a function of voltage changes per minute. It is important to understand that the definition is voltage changes per minute. A pulse would be represented by a negative voltage change and then a positive voltage change. Therefore the double repetition rate has to be taken as

$$\Delta U/\mathrm{min} = 2\times f_{\mathrm{rep}}\ [1/\mathrm{s}]\times 60\ [\mathrm{s/min}]\ . \quad (29)$$

In order to fulfil the required standard a constant power loading for the modulator was developed. The power of the modulator has to be

$$P = U \times I = \mathrm{const}\ ! \quad (30)$$

The current has to be regulated according to Eq. (30), to be the inverse of the $\Delta U$ that occurs during pulsing. If $U$ decreases the current has to be increased by the same ratio.

It has to be taken into account that an SCR bridge generates reactive power dependent on the firing phase angle; therefore, these are not suitable. Instead either a diode bridge rectifier with intermediate DC link or active front-ends with self-commutated converters have to be used.

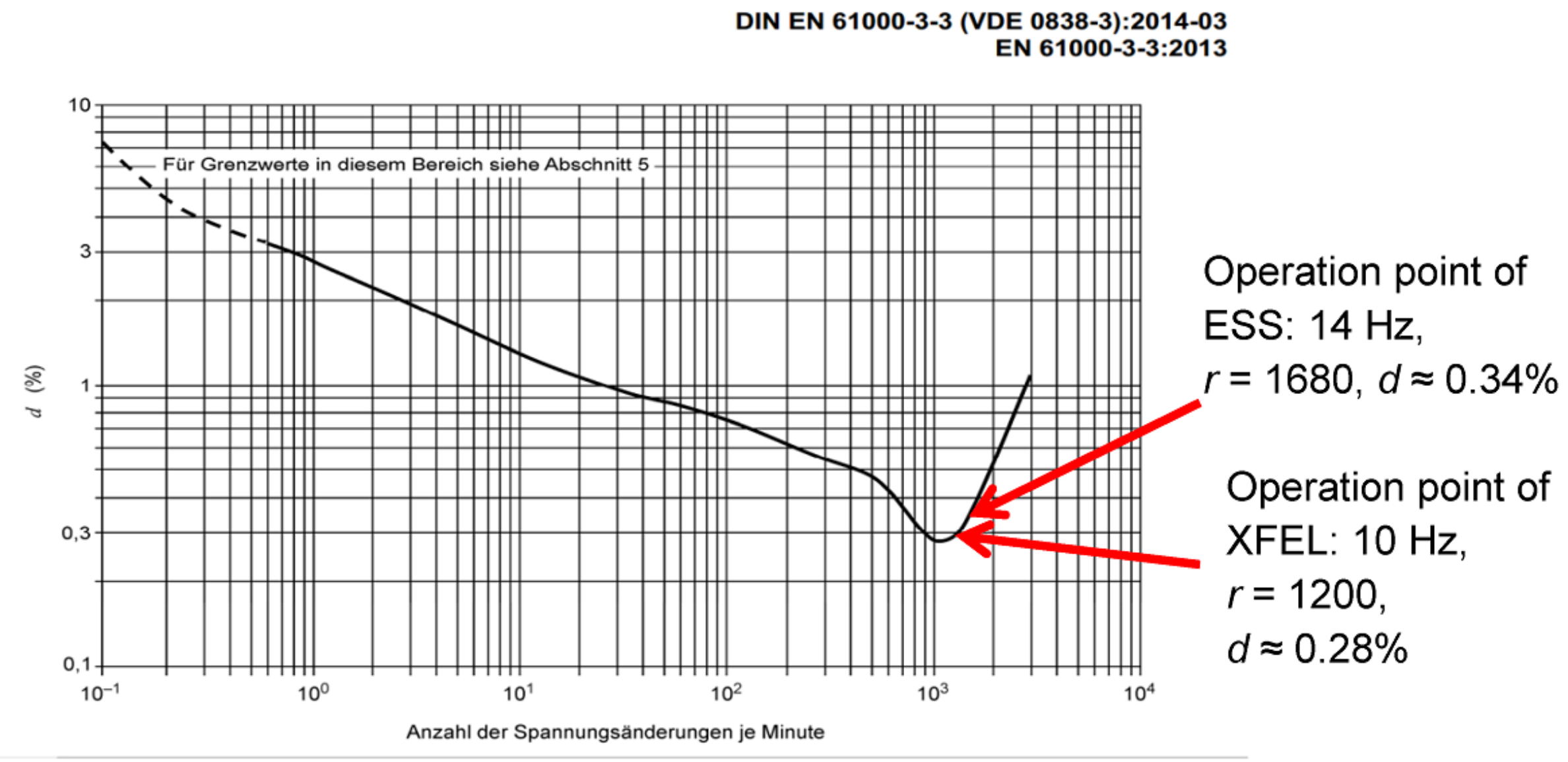

**Fig. 32:** Number of voltage changes per minute with allowed amount of distortion

## 6.1 DESY mains specification and specification of the modulator

Unfortunately, particle accelerators are using repetition rates that are close to the frequency of minimum of allowed distortion. The XFEL is operated at 10 Hz, giving 1200 voltage changes. According to Fig. 32, $d = 0.28\%$. The found value of $d$ defines the actual allowed power changes ($\Delta P$)/modulator. Here, $d$ is defined by the equation

$$d = \frac{\Delta S}{S_{sc}} = \frac{\Delta U}{U_{\mathrm{nom}}} = 0.28\% \qquad (31)$$

where $d$ is allowed distortions, $\Delta S$ is the variation in apparent power, $S_{\mathrm{sc}}$ is the short-circuit power of the mains and $U_{\mathrm{nom}}$ is the nominal voltage of the mains

At DESY the intermediate voltage is 10 kV. The short-circuit power of the mains station to which the modulators are connected to, is 250 MVA. The allowed power variation is:

$$250\ \mathrm{MVA} \times 0.28\% = 700\ \mathrm{kVA}\ . \qquad (32)$$

In the design, including future upgrades, the number of XFEL modulators is set to 35. This would allow 20 kVA/modulator deviations to fulfil the requirement of the grid operator.

Since other components in the machine are assumed to be more critical than the human eye, the budget was cut by two. A value for the specification of 10 kVA/modulator was defined, a challenging, but not impossible value.

## 6.2 Different power supplies for constant power

The first market analysis during the middle of the 1990s showed that no constant-power power supplies were available. Capacitor charging units at that time usually provided the possibility of voltage or constant current regulation. A research and development programme was launched to make a first prototype to cope with this problem. The result was the power supply shown in Fig. 33. This was built as a 300 kW prototype, able to supply a modulator in the test stand [9, 13].

With this topology the converter delivers a constant power into the capacitor $C_{load}$ at a fixed switching frequency. Charging a resonant capacitor C to a *constant voltage* $U_B$ or 0 V, by taking current from the mains, results in a *constant charge* from the mains and is independent of the load.

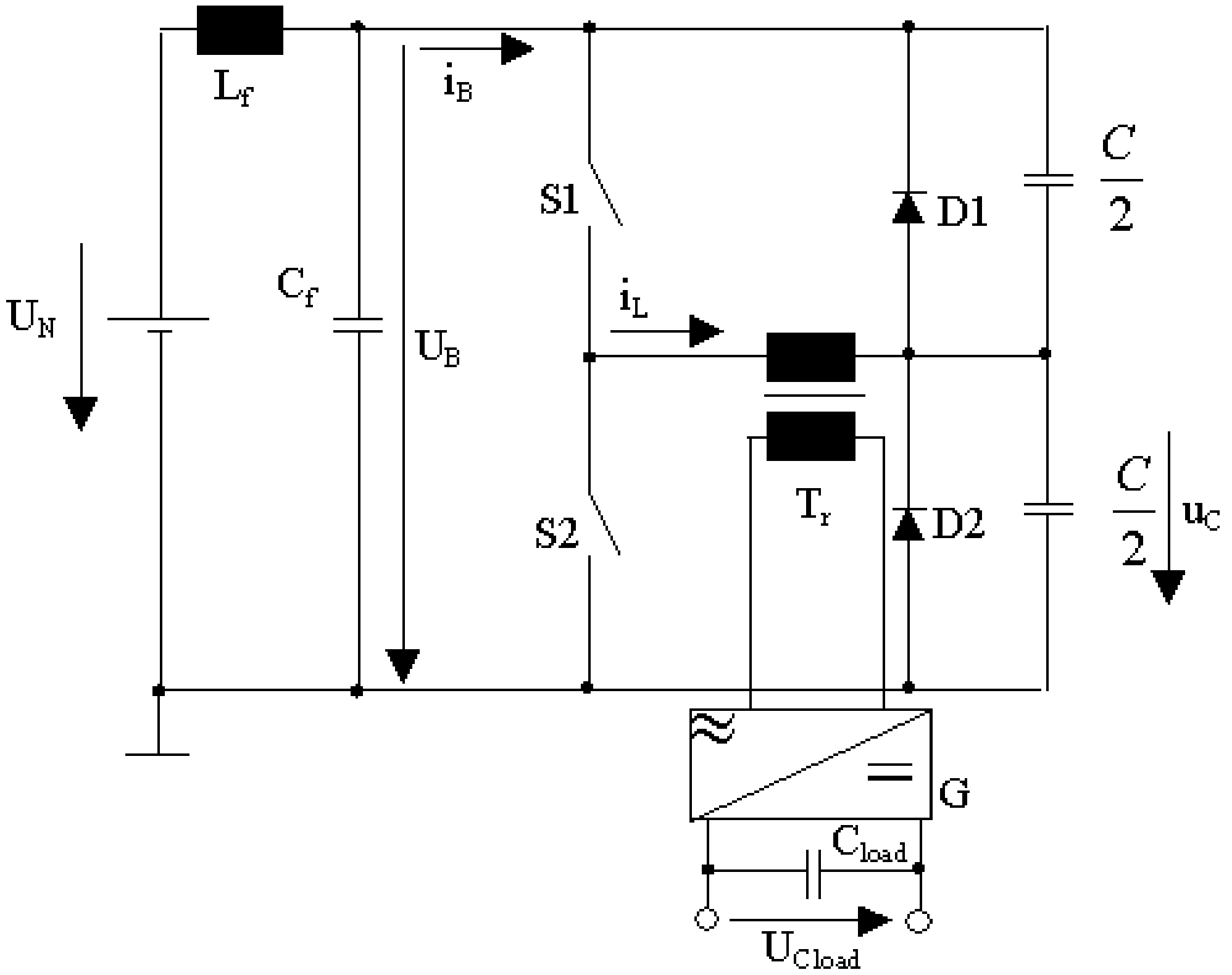

**Fig. 33:** First constant power power supply developed at DESY. G, rectifier; $i_b$, supply current; $i_L$, transformer primary current; $u_C$, resonance capacitor voltage; $U_{Cload}$ output voltage at modulator capacitor bank; $i_{Btl}$, current $i_B$ at time $t_1$; *L*, primary stray inductivity of the transformer; *f*, resonance frequency of the resonant LC circuit; *n*, gear ratio of the transformer and rectifier; *T*, time period of the switching frequency of S1 and S2; C, resonance capacitor; $U_B$, DC link supply voltage; $U_N$, primary rectifier voltage; $C_f$, DC link capacitor; $L_f$, filter inductance.

A very nice analogy of the working principle is the comparison with a water bucket. This bucket is filled with water per period. Afterwards it is poured into a pool. The water quantity that is poured into the pool is dependent upon the filling/pouring frequency. It is independent of the water-level of the pool.

In the tendering process, this topology was not chosen because of the price. Instead, a series combination of buck converters was shown to have a better price. The schematic can be seen in Fig. 34. It was produced by FuG, Germany. These power supplies have been working in FLASH for several years. The regulation is analog and consists of three regulation loops: voltage, current and power regulation.

During normal operation the power is regulated. In the event of too high currents or voltages these regulation loops guarantee that none of the values are exeeded. For example, the power supply stops loading when the required voltage is reached.

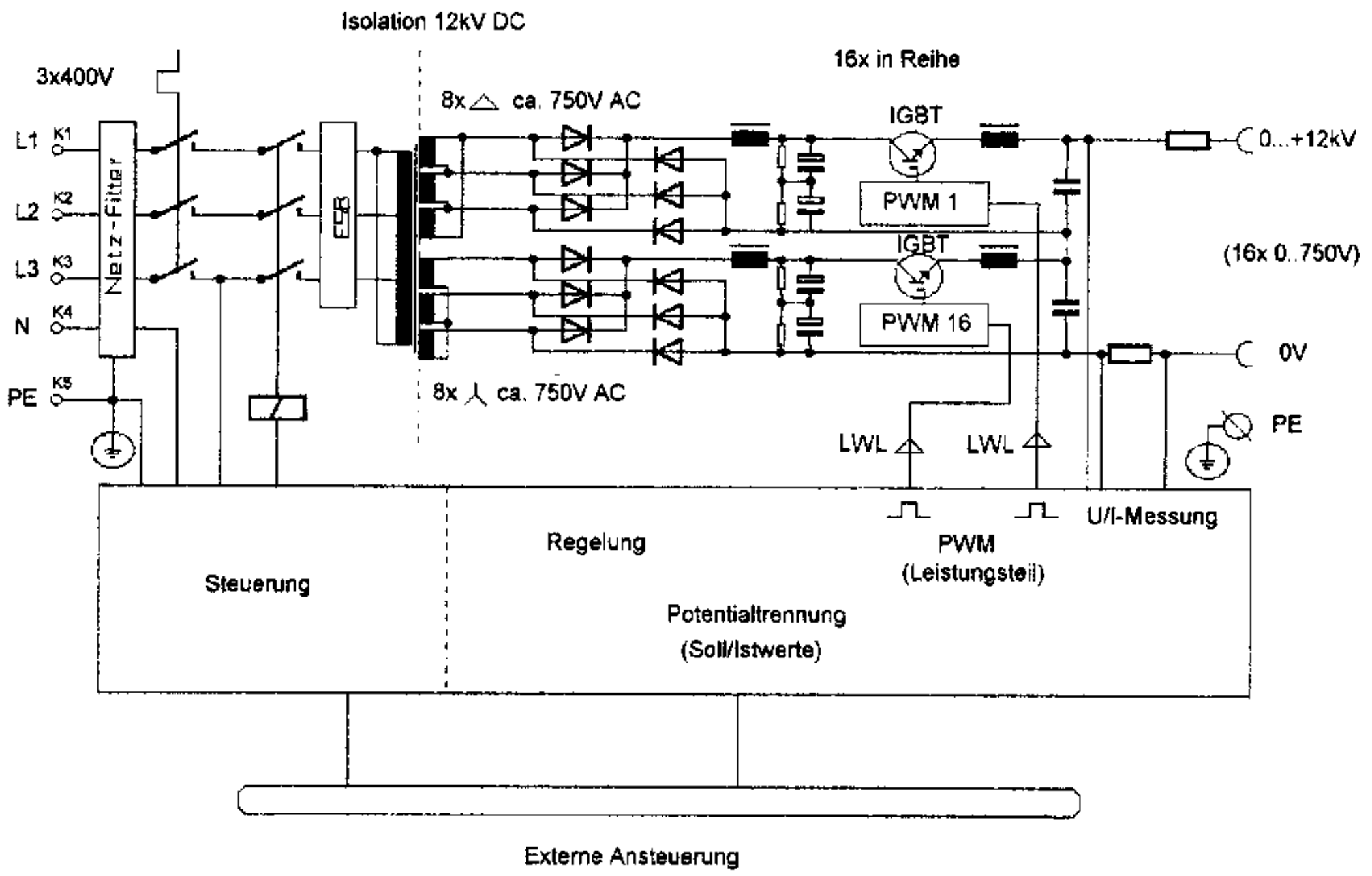

**Fig. 34:** Contant-power power supply using buck converters installed at FLASH

The third solution is installed in the Ampegon modulator for XFEL as shown in Fig. 35. The modules have a combined function of constant power charging (main rectifier and boost converter), energy storage in the capacitor and the pulse-forming IGBT. The boost converter is regulated in constant power mode according to Eq. (30).

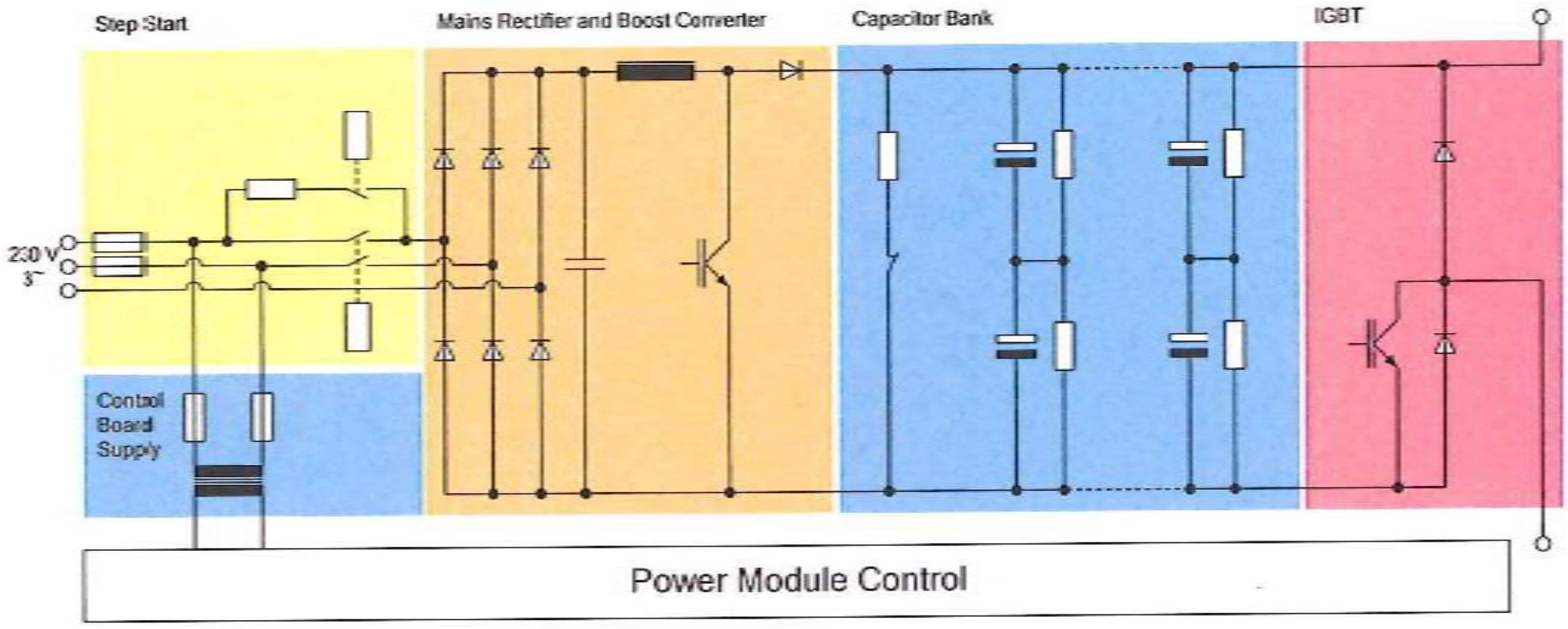

**Fig. 35:** Ampegon power module, combining constant power loading, energy storage and pulse IGBT

The measurement results are shown in Figs. 36 and 37. Figure 36 shows the waveform that was measured during the acceptance test. The 50 Hz fundamental wave is taken to calculate power deviation due to pulsing.

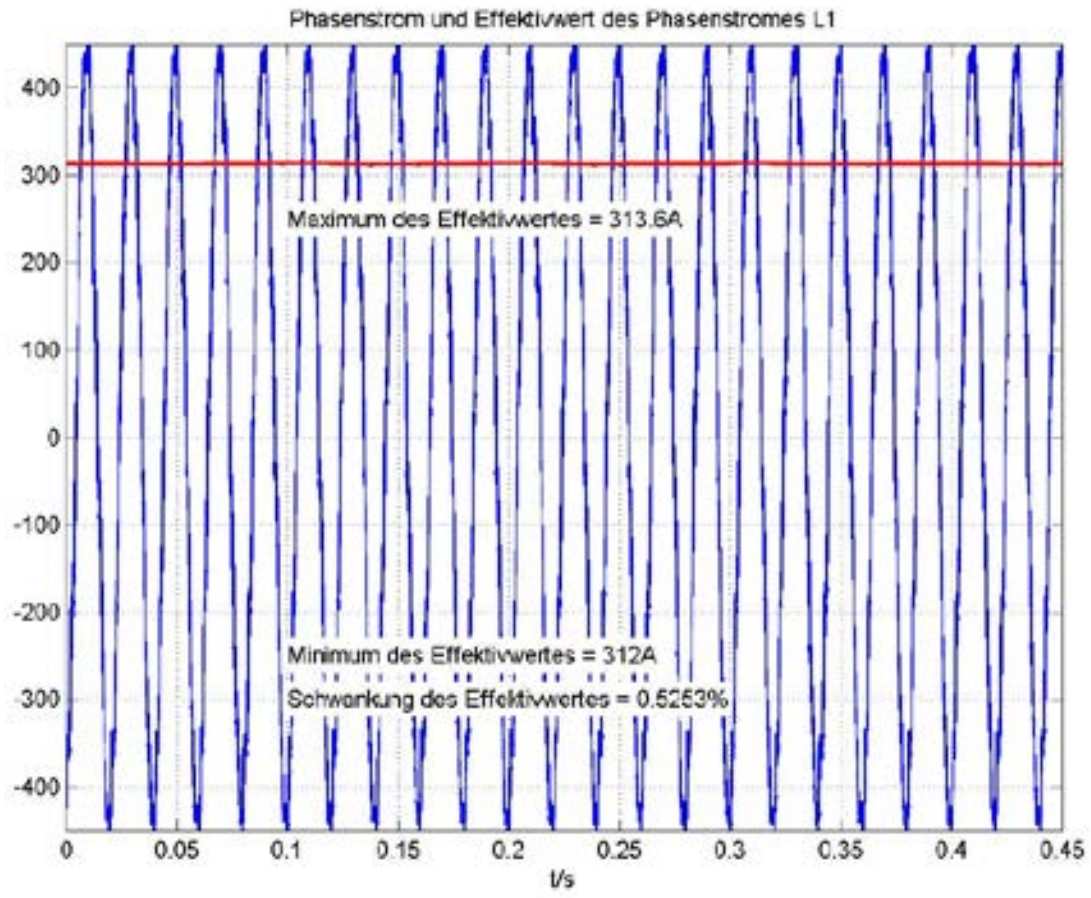

**Fig. 36:** 50 Hz signal with measured 10 Hz deviation

As described the specified value of power variations should be <10 kVA/modulator. This would correspond to a current variation of

$$\Delta I = \frac{\Delta S}{\sqrt{3} \times U} = \frac{10\ \mathrm{kVA}}{\sqrt{3} \times 690\ \mathrm{V}} = 8.4\ \mathrm{A}\ . \tag{33}$$

The measured result shows that:

$$\Delta I = 2.5\ \mathrm{A}\ \text{what corresponds to a}\ \ \Delta S \approx 3\ \mathrm{kVA}\ . \tag{34}$$

The performance of the power regulation is approximately a factor of 3 better than the requirement. Figure 37 shows the measured three-phase currents of a modulator input. The typical frequencies of the diode rectifier also appear. However, no variations in the peak-to-peak values of the amplitude or frequency variations occur.

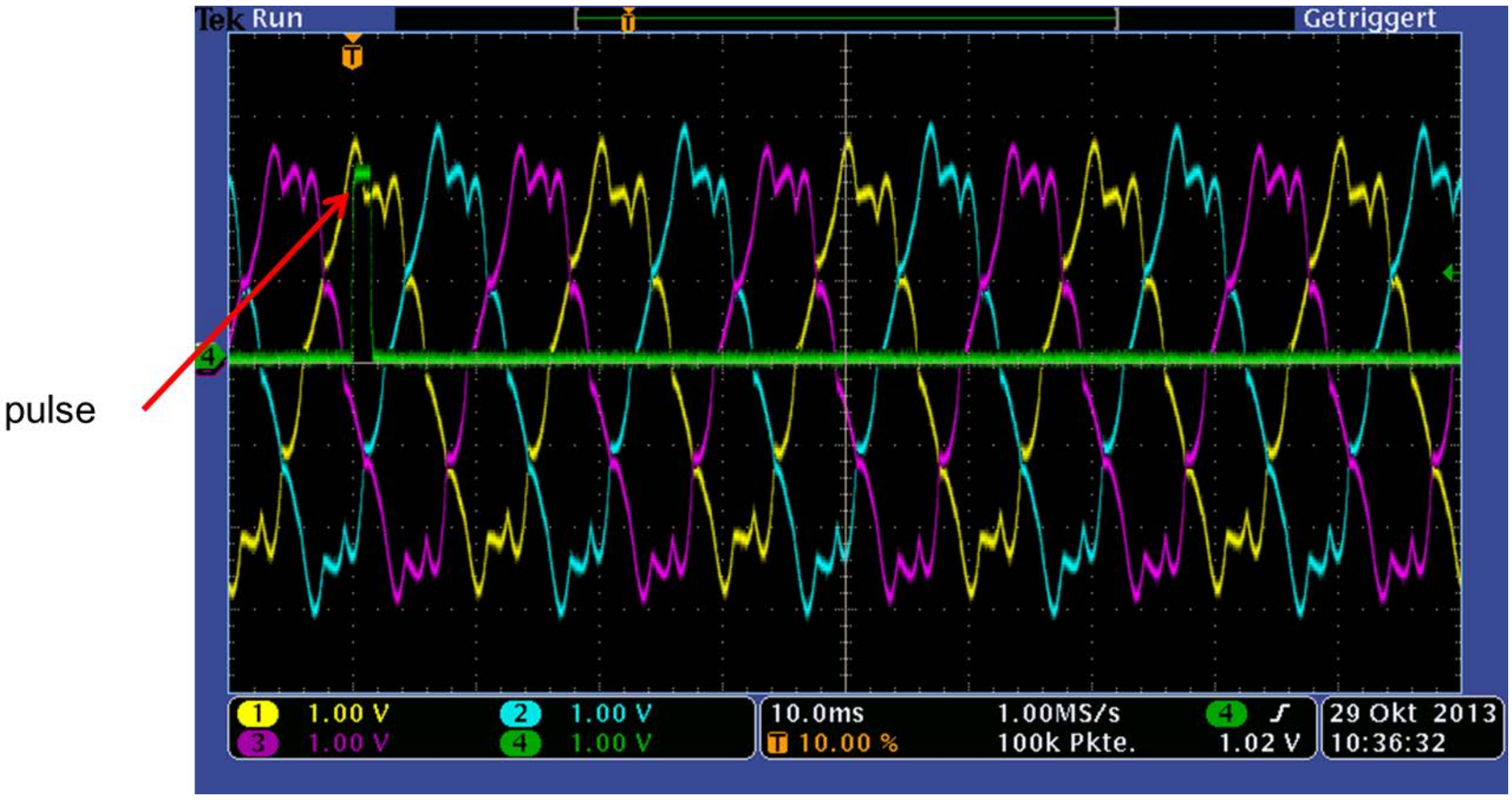

**Fig. 37:** Measured input current; waveform is not influenced by the modulator pulse. Yellow, light blue, pink: line currents of modulator (270A). Green: HV-pulse (9kV).

## 7 EMI aspects

In the pulsed modulator applications one deals with high voltage and high currents that are switched at high d*V*/d*t* and d*i*/d*t* values. It is obvious that it is extremely important to start, from the very beginning, building in features that minimize EMI.

Presenting a complete overview of correct EMI construction in this paper is not possible. Only a few very basic hints are given:

- think in currents instead of voltages: find the paths where the return current will flow;
- never think that a cable is just a conductor, it is also an inductance;
- each component can and will form a resonant circuit with its neighbour;
- put the current path and return path close together;
- use low inductive ground planes;
- read, study and put into practice the lectures by A. Charoy during this CAS and [15].

In Fig. 38 the Ampegon modulator is shown. A few additional stray components are inserted into the basic schematic. These are capacitors to ground that achieve a high d*V*/d*t* and inductances in the return paths of the high currents.

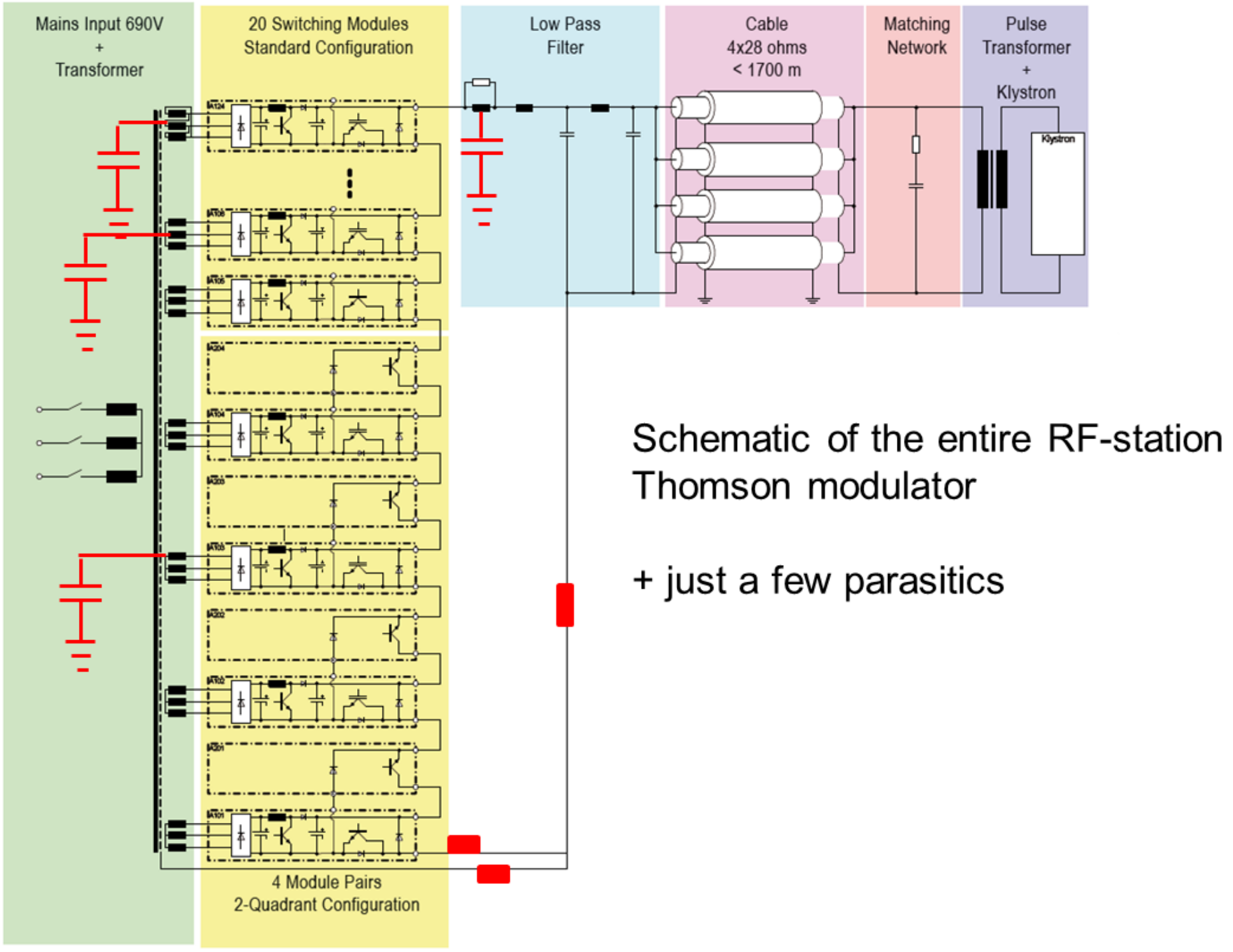

**Fig. 38:** Ampegon modulator with a few parasitics

Figure 39 is a simulation model of the PPT modulator for FLASH in combination with the pulse cables [14]. The circle indicates inductance L3 in the return path. The d*i*/d*t* of the main pulse is transformed into a real voltage change. The inductance represents a voltage source that produces common mode currents. In Fig. 39 every component that is connected to this inductance experiences a severe change of ground potential.

Please remember: minimize the inductances in the current's return path!

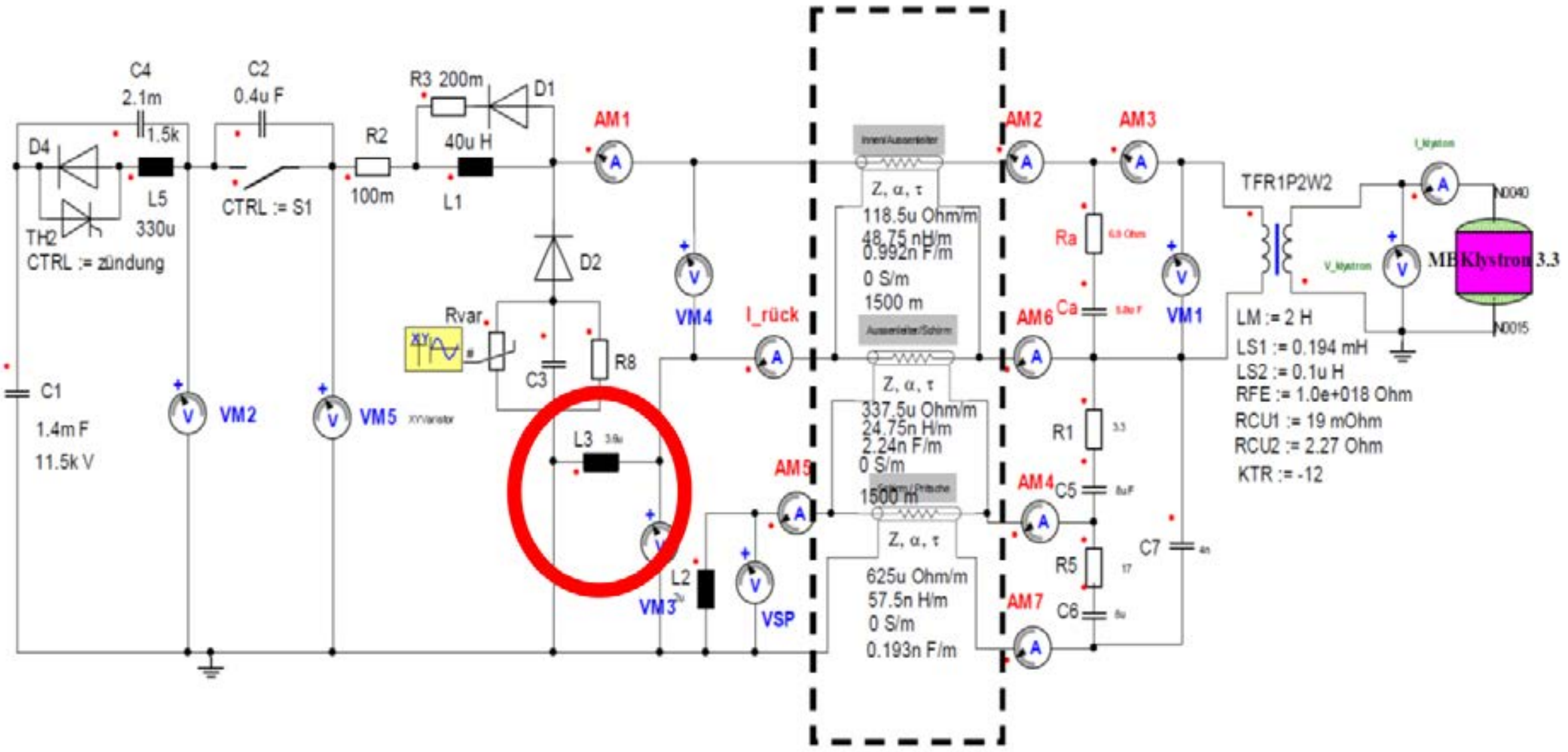

**Fig. 39:** Simulation model of the bouncer modulator. L3 is common mode relevant

## 8 Future developments

The ongoing development of semiconductors and new materials for magnetic devices allow new approaches. High switching frequencies of semiconductors are possible, and transformers are able to handles them without large losses. In general, there is a trend to modules with lower voltage at switching levels. By this the series connection of large semiconductor devices is obsolete. The large pulse transformer will be replaced by smaller high-frequency types. The advantage is that less energy is stored during the pulse, which makes klystron protection much easier.

Three types of modulators will be introduced. These are either under construction or prototypes are being built.

### 8.1 JEMA hybrid inverter Marx system with custom potted transformer

The first modulator is presented by JEMA [5], and is shown in Fig. 40. It is called the hybrid inverter Marx system with custom potted transformers. A primary SCR rectifier supplies a common DC bus. The SCR rectifier is only used for the soft start and loading the intermediate DC bus. Once it is loaded, the bridge is in permanent conduction mode and reacts as a diode bridge. Paralleled H-bridges installed in modules deliver the required AC waveformThe transformer consists of a primary winding and several secondary windings. . By stacking thesecondary windiings , the output HV is achieved. The switching frequency is 4 kHz. The flicker compensation in this proposal is done via the DC link reactor. The inductance would be large for small repetition rates and would have to be looked at in detail when the requirements are known.

### 8.2 The stacked multi-level topology

The stacked multi-level (SML) topology in Fig. 41 was proposed by C. Martins (ESS). It is in the prototyping phase in collaboration with the University of Lund.

It also works with several modules at the 1 kV level with IGBTs as semiconductor switches, operating at 16 kHz. Each module has its own transformer and rectifying unit. Several of these units are connected in series at the output, forming the HV pulse.

Each module comprises power factor correction and precise capacitor charging. Here again, constant power charging is done to reduce flicker. Energy is stored in a capacitor. An H-bridge excites a resonant circuit to supply the transformer. At the secondary winding of the transformer a rectifier with LC filter is introduced.

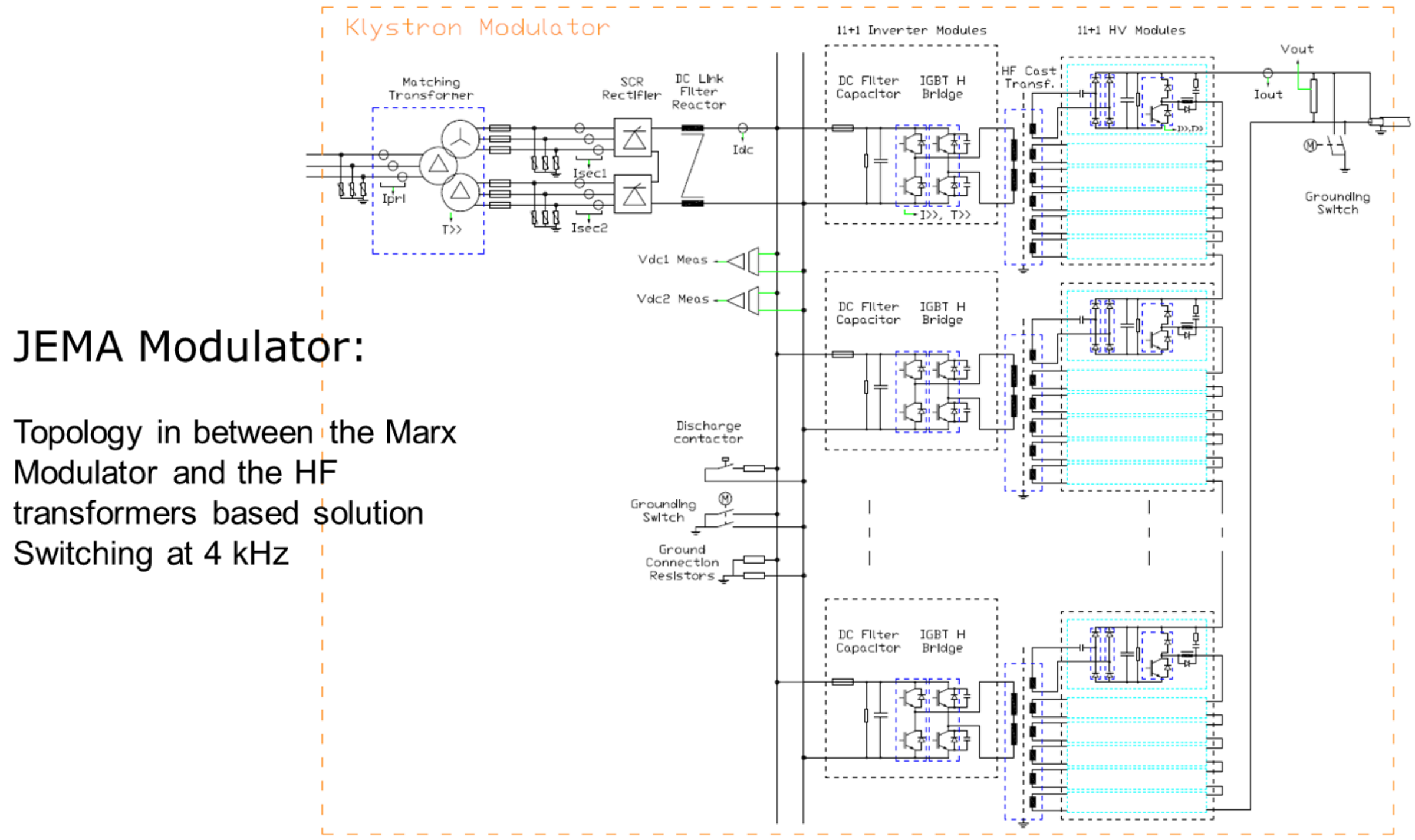

**Fig. 40:** Hybrid inverter Marx system with custom potted transformers

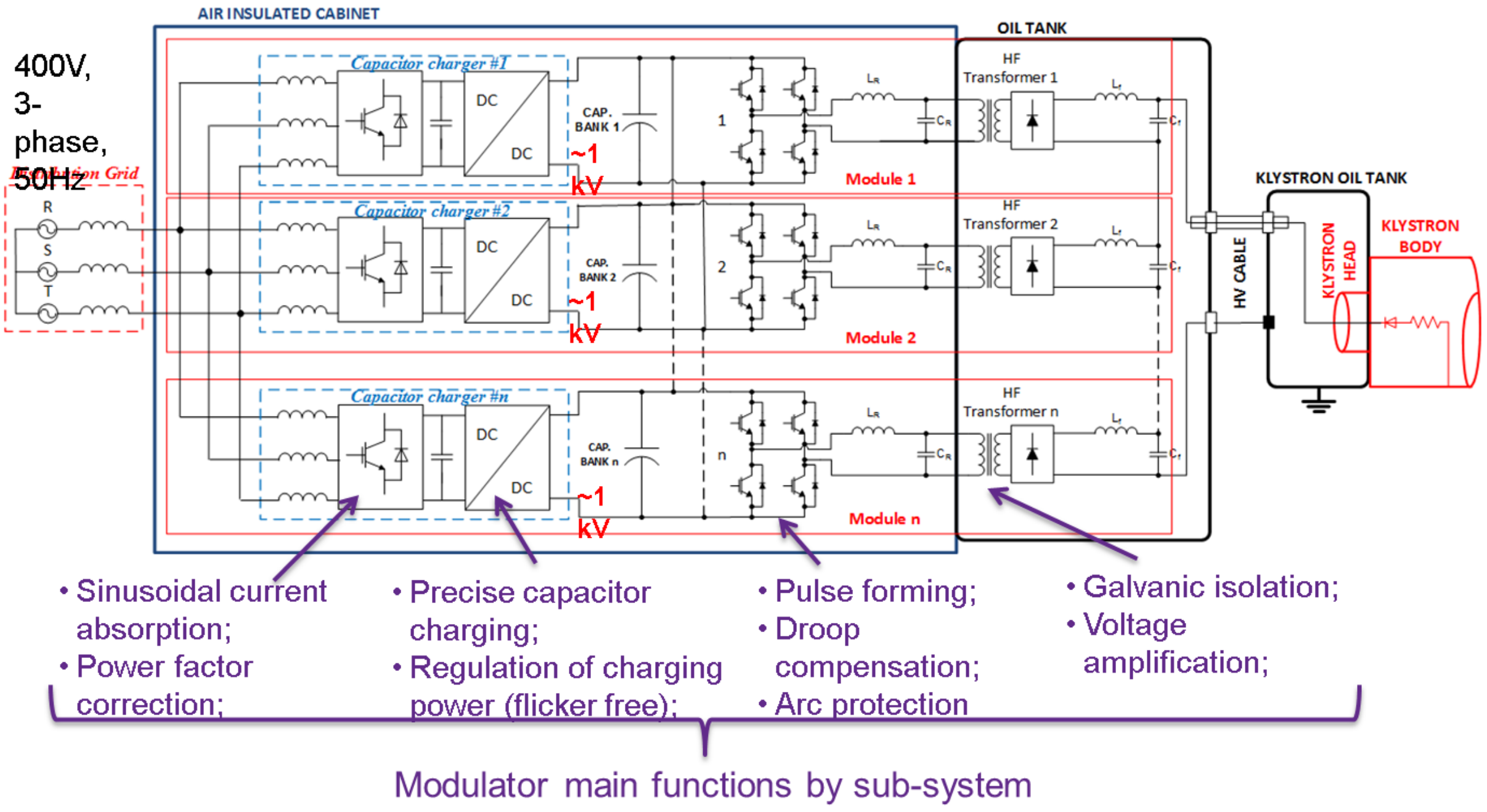

**Fig. 41:** Stacked multi-level (SML) topology, ESS

## 8.3 Ampegon proposal for ESS modulator

The last proposal is introduced by Ampegon for ESS. The basic topology of the schematic shown in Fig. 42 is very similar to the ESS SML topology. However, there are different approaches. The intermediate voltage is at 400 V. Instead of using IGBTs, metal oxide semiconductor field-effect transistors (MOSFETs) will be used, allowing a switching frequency of up to 150 kHz. This high switching frequency allows a very compact transformer, as also shown in Fig. 42.

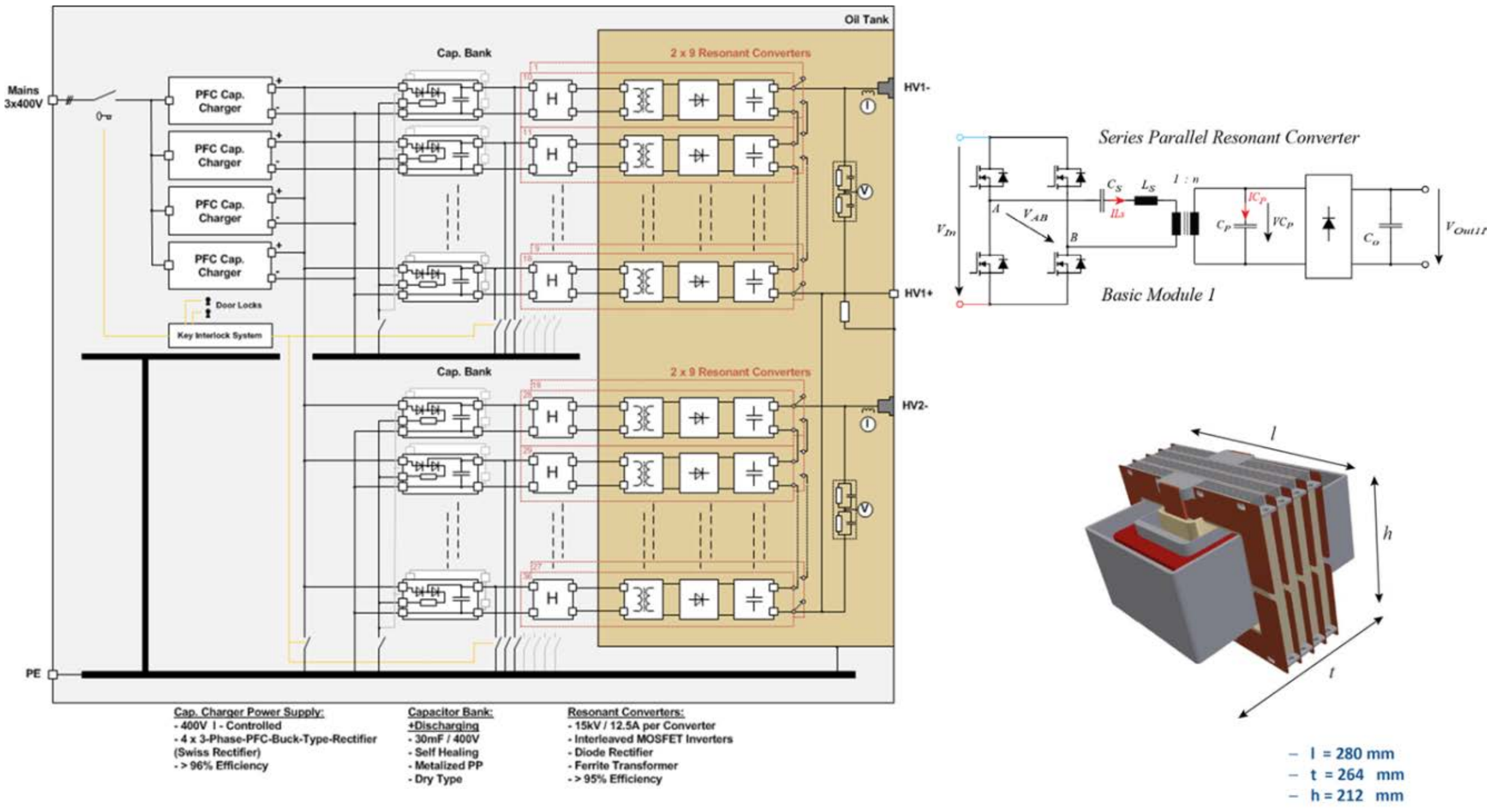

**Fig. 42:** Ampegon proposal for the ESS modulator with new transformer

## 8.4 New semiconductor technologies

Although beneficial for high voltages and fast switching, components made of silicon carbide or gallium arsenide are not yet in use. There is additional potential here for future applications.

# 9 Conclusions

It has been an interesting time for developments and improvements in long pulse modulators. This has been from the first idea, to prototyping, and finally to the series production of different topologies. The questions that had to be solved, ranging from the basic pulse-forming units, klystron protection and flicker rejection with a constant power loading, were solved at different accelerators. However, the improvements achieved by new technologies are still ongoing.

In the near future several large projects will use long pulse modulators. Some of these are:

– XFEL commisioning and operation;

– European Spallation Source (ESS);

– International Linear Collider (ILC);

– Project X;

– CLIC.

With the construction of modulators for these machines, power electronic engineers will have an interesting time and a lot of fun in the near future.